# INTERACTIONS, SYMMETRY BREAKING, AND EFFECTIVE FIELDS FROM QUARKS TO NUCLEI
## (A primer in nuclear theory)

JACEK DOBACZEWSKI

*Institute of Theoretical Physics, Warsaw University*
*Hoża 69, PL-00681, Warsaw, Poland*

**Résumé**

Ce cours présente une introduction à la théorie des systèmes nucléaires partant des champs de quarks et de gluons tels qu'ils sont décrits dans la chromodynamique quantique puis discutant les propriétés des mésons $\pi$ et des nucléons, les interactions entre nucléons et la structure du deuteron et des noyaux légers pour arriver à la description des noyaux lourds. Ceci montre comment notre description des systèmes nucléaires dépend des différentes échelles d'énergie et de distance et des concepts de champ effectif et de brisure de symétries.

**Abstract**

An introduction to nuclear theory is given starting from the quantum chromodynamics foundations for quark and gluon fields, then discussing properties of pions and nucleons, interactions between nucleons, structure of the deuteron and light nuclei, and finishing at the description of heavy nuclei. It is shown how concepts of different energy and size scales and ideas related to effective fields and symmetry breaking, enter our description of nuclear systems.

## 1 INTRODUCTION

Modern nuclear physics contains a much larger class of subdomains than it did even only ten years ago. The change in the meaning of the name reflects changes in physicists' minds. It is now widely recognized that there exist a unity in the way we perceive all physical systems for which the quantum chromodynamics (QCD) is a fundamental theory. This embraces the QCD vacuum, quarks and gluons, composite particles, like pions and nucleons, and nuclei as aggregates of nucleons.

Typical scales of energy and size range here from 1 GeV to 1 keV, and from 0.1 fm to 10 fm, but tools and methods that are used to describe all these systems are very much alike. In particular, in order to cope with difficulties related to the complication of structure of these systems, one has to invoke ideas of the effective field theory (EFT), which separate our approach into several stages of description. Although links between these stages cannot be attacked, at present, with exact methods, at every one of them we can obtain successful understanding of the physical reality. Moreover, methods based on the concept of symmetry breaking are by now standard throughout the domain.

Certainly, the nuclear physics, in this larger sense, is far too broad a domain for a single physicist, and we are forced to specialize in much narrower subfields. However, it is essential that we learn enough of the whole of it, in order to be able to communicate and understand one another. These lectures are prepared with such a goal in mind.

Nuclear physics in three lectures might seem to be an impossible task, and of course it is. There is no point in attempting a balanced or representative overview of neither facts nor approaches. The choices I made below are highly personal; I have tried to discuss things that show similarities of different aspects of the field, and a general philosophy of how we do the business.

Of course, the main question is from where to start such lectures. The background that students carry out from undergraduate and graduate courses differs very much from country to country, and from university to university, and is often meagre. Even worse, students are often told that they can "understand" physics without actually learning it. I know, learning is a painful process and intelligent human beings request being liberated from this pain – then they become not physicists but lawyers. In physics, in my opinion, there is no understanding without learning. On the other hand, neither there is learning without teaching, so my first task here is to teach you things that you need to know to follow the course.

The first part of the course (I called it the first four minutes) gives you an overview of elements that are profusely used in the following. It is meant to give you the list of things, and references to main textbooks, rather than real knowledge – each minute here is usually taught one semester at the university. However, there is no understanding of the micro-world without at least two basic abilities: one has to know how to read a Lagrangian and one has to know how to use creation and annihilation operators. This is the mother tongue, which you have to learn as apprentice in nuclear physics.

## 2 QUANTUM FIELDS OF NUCLEAR SYSTEMS

### 2.1 Quantum Field Theory in Four Minutes

#### 2.1.1 Minute No. 1, the Classical Mechanics

Classical systems [1] are described by defining two elements: 1° – the set of classical coordinates $q_i$, which are supposed to give a complete information about the state of the system, and 2° – the Lagrangian. The state depends on a parameter called the classical absolute time $t$, and hence, coordinates $q_i(t)$ are functions of time. The Lagrangian,

$$L = L(q_i, \dot{q}_i, t) = T - U, \tag{1}$$

is a function of coordinates $q_i$, velocities $\dot{q}_i$, and time $t$. According to the mechanistic point of view of the classical mechanics, every system in our Universe, including the whole Universe, is fully described by finding its coordinates and Lagrangian. For most systems the Lagrangian is equal to a difference of the kinetic energy $T$, depending only on velocities, and the potential energy $U$, depending only on coordinates [see the second member of Eq. (1)].

Once the system is defined as above, its properties can be derived from simple principles. The time evolution of the system can be found from the principle of extremal action $I$,

$$\delta I = \delta \int_{t_1}^{t_2} \mathrm{d}t L(q_i, \dot{q}_i, t) = 0, \tag{2}$$

which gives the Euler-Lagrange equations,

$$\frac{d}{dt}\frac{\partial L}{\partial \dot{q}_i} - \frac{\partial L}{\partial q_i} = 0. \tag{3}$$

This leads to a set coupled differential equations that can be, in principle, solved once the initial conditions $q_i(t{=}0)$ and $\dot{q}_i(t{=}0)$ are known. One thus obtains the complete past and future history of the system $q_i(t)$. The rest is just a technicality ;) of how to solve differential equations. For typical systems, the kinetic energy is a quadratic function of velocities, for which the Euler-Lagrange equations are linear – and can be solved fairly easily.

Although we do not really need it in classical mechanics, we shell also introduce the formulation in terms of the Hamiltonian $H$. This gives us a bridge towards the quantum mechanics. Namely, we define the classical momentum $p_i$ by

$$p_i = \frac{\partial L}{\partial \dot{q}_i}, \tag{4}$$

and we transform the Lagrangian into the Hamiltonian,

$$H(q_i, p_i, t) = \sum_i p_i \dot{q}_i - L, \tag{5}$$

as well as the Euler-Lagrange equations into the Hamilton equations,

$$\frac{\partial H}{\partial p_i} = \dot{q}_i, \quad \frac{\partial H}{\partial q_i} = -\dot{p}_i. \tag{6}$$

### 2.1.2 Minute No. 2, the Quantum Mechanics

Quantum systems [2] are described by the wave function $\Psi(q_i, t)$ (complex function of coordinates $q_i$ and time $t$), and by the Hamilton operator $\hat{H}$ that can be obtained from the classical Hamiltonian by a procedure called quantization. We define operators that correspond to each classical object, e.g., the classical coordinates and momenta are quantized as,

$$q_i \longrightarrow \hat{q}_i = q_i, \quad p_i \longrightarrow \hat{p}_i = \frac{\partial}{\partial q_i}. \tag{7}$$

Then, the Hamilton operator is, more or less, obtained by inserting these operators into the classical Hamiltonian, i.e.,

$$H(q_i, p_i, t) \longrightarrow \hat{H}(\hat{q}_i, \hat{p}_i, t). \tag{8}$$

This is not an exact science, because the function of operators cannot be uniquely defined for a given function of variables; one has to also define the order in which the operators act. Well, in fact the quantization provides us only with general rules on how to start the quantum mechanics based on our knowledge of the classical mechanics. One can also subscribe to the point of view that we must axiomatically define the quantum system by specifying its Hamilton operator. Once this is done, the time evolution of the system (of its wave function) is given by the Schrödinger equation,

$$i\hbar \frac{\partial}{\partial t} \Psi(q_i, t) = \hat{H}(\hat{q}_i, \hat{p}_i, t) \Psi(q_i, t). \tag{9}$$

This leads to a set coupled differential equations that can be, in principle, solved once the initial conditions $\Psi(q_i, t{=}0)$ are known. One thus obtains the complete past and future history of the system $\Psi(q_i, t)$. The rest is just a technicality ;) of how to solve differential equations.

Quantum mechanics also adds a pivotal element to our understanding of how our world works, namely, the probabilistic interpretation. In classical mechanics, once our Euler-Lagrange equations give us the set of coordinates $q_i$ at time $t$, the experiment performed at time $t$ to find the system at point $q_i$ yields one possible answer: the system is there. In quantum mechanics, the same experiment yields the answer that the system is within volume $dV$ around $q_i$ with probability $|\Psi(q_i,t)|^2 dV$ and the answer that it is not there, with probability $1 - |\Psi(q_i,t)|^2 dV$. Hélas, it seems that the world is just like that, nothing is certain any more. However, at least the probabilities of obtaining given experimental answers can be rigorously calculated.

### 2.1.3 Minute No. 3, the Classical Field Theory

The classical field theory [3] describes certain physical systems as infinite-dimensional classical objects whose states need as many classical coordinates as there are points in the 3D space. Therefore, index $i$ that two minutes ago was used to enumerate the classical coordinates, now changes into the space point $\boldsymbol{r}$, and the coordinate itself – into the value of a certain function $\psi(\boldsymbol{r})$, called the field, at point $\boldsymbol{r}$,

$$i \longrightarrow \boldsymbol{r}, \quad q_i \longrightarrow \psi_{\boldsymbol{r}} \equiv \psi(\boldsymbol{r}). \tag{10}$$

Local Lagrangian density $\mathcal{L}[\psi(\boldsymbol{r}), \partial_\mu \psi(\boldsymbol{r})]$ defines the Lagrangian,

$$L = \int d^3\boldsymbol{r} \; \mathcal{L}[\psi(\boldsymbol{r}), \boldsymbol{\nabla}\psi(\boldsymbol{r})], \tag{11}$$

and the extremal-action principle (2) gives the same Euler-Lagrange equations, which are now called field equations. Since the Lagrangian now depends on spatial derivatives of fields, the field equations are differential equations both in time and space. They can be, in principle, solved once the initial fields $\psi(\boldsymbol{r},t=0)$ and spatial boundary conditions $\psi(\boldsymbol{r} \in \text{border}, t)$ are known. One thus obtains the complete past and future history of the system $\psi(\boldsymbol{r},t)$. The rest is just a technicality ;) of how to solve differential equations.

In the physical world, the classical fields described above replace forces that act between particles. The whole Universe is thus composed of (classical) particles and (classical) fields. Particles are sources of fields, and fields exert forces on particles. The novelty here is the notion that a particle does not "feel" other particles; it only feels the fields generated by other particles. The so-called action at a distance disappeared from the theory, and a change of position of one particle, influences other particles only after the field it generates propagates to the rest of the world.

It is clear that the classical field theory is tailored to address the question of time propagation of fields, and makes the full sense within the relativistic approach where all fields propagate with one common and unchangeable velocity. Classical electrodynamics and classical gravity are theories of this type. Relativistic invariance takes then the place of a basic ingredient of the theory, and, e.g., action corresponding to Lagrangian (11) is manifestly relativistically invariant,

$$I = \int dt \; L[\psi(\boldsymbol{r}), \dot{\psi}(\boldsymbol{r})] = \int d^4x \; \mathcal{L}[\psi(x), \partial_\mu \psi(x)], \tag{12}$$

because the four-dimensional volume element $d^4x$ is relativistically invariant. Here we introduced the standard four-vector notation of $x^\mu \equiv (t, \boldsymbol{r})$ and $\partial_\mu \equiv \partial/\partial x^\mu$.

### 2.1.4 Minute No. 4, the Quantum Field Theory

Quantum field theory [4] performs quantization of classical fields in a very much the same way as the quantum mechanics performs quantization of classical coordinates. The field wave function now becomes a functional $\Psi[\psi(x)]$ of the field $\psi(x)$, and the quantum fields and the quantum conjugate momenta are

$$\hat{\psi}(x) = \psi(x) \quad , \quad \widehat{\text{momentum}} = \frac{\delta}{\delta \psi(x)}, \tag{13}$$

where $\delta$ denotes the functional derivative. The Schrödinger equation (9) now becomes the set of infinite number of differential equations – a pretty complicated thing. I somehow hesitate to write that the rest is just a technicality of how to solve it. In principle, nothing special has happened. The same rules have been applied and an analogous, albeit much more complicated, set of equations emerged. However, we are very, very far from even approaching a possibility of exact solutions of this set. We are not at all going to embark on discussing these questions here. Basic physical picture of the quantum field theory can be very well discussed in terms of its classical counterpart, and in terms of classical-field Lagrangian densities discussed during the third minute above. It as amazing how much can be said about properties of the micro-world by just specifying what are the symmetries and the basic couplings between the classical fields. Below we follow this way of presenting properties of strongly interacting systems.

The new, qualitatively different, element introduced by the quantum field theory is that particles now disappeared from our description of the physical world – there are only fields. One does not distinguish which is the object that "exists" and which is the object that "transmits forces". All fields have both these characteristics simultaneously; which field interacts with which, and in which way, is fully specified by the Lagrangian density.

## 2.2 Quantum Electrodynamics (QED)

Classical [5] and quantum [6] electrodynamics are probably the best established theories of our world. They describe interactions between charged objects, where by the charge we mean the traditional electric charge. Quantum electrodynamics (QED) allows to calculate electrodynamic properties of particles to an unbelievable precision, e.g., the magnetic moment of the electron, calculated up to the eighth order of the perturbation theory [7], and the measured value [8],

$$\mu_e^{\text{th}} = 1.001\ 159\ 652\ 153\ 5(280)(12) \tfrac{e\hbar}{2mc}, \tag{14}$$
$$\mu_e^{\text{ex}} = 1.001\ 159\ 652\ 188\ 4(43) \tfrac{e\hbar}{2mc}, \tag{15}$$

are in excellent agreement. Moreover, the error of the theoretical value comes mostly from the uncertainty in the measured value of the fine structure constant $\alpha$ (the first error), and less from estimated higher-order effects (the second error).

For an electron coupled to the electromagnetic field, the Lagrangian density, from which everything can be derived, reads

$$\mathcal{L} = -\tfrac{1}{4} F_{\mu\nu} F^{\mu\nu} - \bar{\psi}_e \gamma^\mu [\partial_\mu + ieA_\mu] \psi_e - m_e \bar{\psi}_e \psi_e. \tag{16}$$

It is expressed within the relativistic formalism that uses space-time four-coordinates numbered by indices $\mu, \nu$=0,1,2,3. Moreover, we assume that each pair of repeated indices implies summation over them. Here and bellow we use the units defined by $\hbar = c = 1$, for which the elementary charge, $e = \sqrt{4\pi\alpha}$, is a dimensionless quantity depending on the fine-structure constant $1/\alpha \simeq 137$. (Note that the elementary charge $e$ is positive, while the charge of the electron

$q=-e$ is negative.) In such a unit system, the only unit left is the energy, so for example, the momentum has the unit of energy, position and time – the unit of (energy)$^{-1}$, and the Lagrangian density $[\mathcal{L}]$=(energy)$^4$ (when $\mathcal{L}$ is integrated over the space-time it gives the dimensionless action $I$).

The first term in the QED Lagrangian density (16) describes the free electromagnetic field defined by the four-potential $A^\mu \equiv (\phi, \boldsymbol{A})$, containing the standard scalar (Coulomb) potential $\phi$ and vector potential $\boldsymbol{A}$. The electromagnetic field tensor $F_{\mu\nu}$ is defined as

$$F_{\mu\nu} = \partial_\mu A_\nu - \partial_\nu A_\mu. \tag{17}$$

The Euler-Lagrange equations corresponding to this term give the Maxwell equations in free space, i.e., all properties of electromagnetic waves.

The last term in (16) describes the free electron of mass $m_e$ at rest. Its field $\psi_e$ has the structure of the four-component Dirac spinor, but traditionally we do not explicitly show in Lagrangian densities the corresponding indices. The first member of the middle term (the one with $\partial_\mu$) describes the kinetic energy of the electron, and together with the mass term, they constitute the Lagrangian density of a free electron. The corresponding Euler-Lagrange equations give the Dirac equation, i.e., all plane-wave propagation of an electron (and positron) in an otherwise empty space. The kinetic term contains the Dirac 4×4 matrices $\gamma^\mu$ defined by

$$\gamma^0 = -i \begin{pmatrix} 0 & 1 \\ 1 & 0 \end{pmatrix}, \quad \boldsymbol{\gamma} = -i \begin{pmatrix} 0 & \boldsymbol{\sigma} \\ -\boldsymbol{\sigma} & 0 \end{pmatrix}, \quad \gamma_5 = \begin{pmatrix} 1 & 0 \\ 0 & -1 \end{pmatrix}, \tag{18}$$

where the standard Pauli 2×2 matrices $\boldsymbol{\sigma}$ read

$$\sigma_1 = \begin{pmatrix} 0 & 1 \\ 1 & 0 \end{pmatrix}, \quad \sigma_2 = \begin{pmatrix} 0 & -i \\ i & 0 \end{pmatrix}, \quad \sigma_3 = \begin{pmatrix} 1 & 0 \\ 0 & -1 \end{pmatrix}. \tag{19}$$

Finally, the second member of the middle term in (16) describes interaction of the electron with the electromagnetic field. On the one hand, when considered together with the free-electron Lagrangian it gives the Lorentz force that acts on the electromagnetic four-current of the electron,

$$J^\mu = ie\bar{\psi}_e \gamma^\mu \psi_e. \tag{20}$$

On the other hand, when considered together with the free-electromagnetic-field Lagrangian, it gives the source terms in the Maxwell equations that correspond to the same electron current $J^\mu$. The structure of the middle term is dictated by the local gauge invariance of the QED Lagrangian density, i.e., invariance with respect to multiplying the electron field by a position-dependent phase. Such a local gauge invariance is at the heart of constructing the Lagrangian densities for all quantum-field theories applicable to the real world. We shall not discuss these aspects during the present course.

Although we only verbally described the role of each term in the QED Lagrangian density (16), derivation and application of the Euler-Lagrange equations is a standard route. However long, painful, and complicated this route might be, it is a well-paved and marked way to get physical answers. In practice, it has already been followed way up, towards incredibly remote summits.

## 2.3 Quantum Chromodynamics (QCD)

It is remarkable that quantum chromodynamics [9] that describes all phenomena related to strongly interacting particles, can be constructed in full analogy to the QED. The "only" difference

Figure 1: Fermion building blocks for electroweak (left) and strong (right) interactions. From `http://www.cpepweb.org/`.

is a more complicated gauge group, SU(3) instead of U(1). The complete QCD Lagrangian density reads

$$\mathcal{L} = -\tfrac{1}{4}F^\alpha_{\mu\nu}F^{\mu\nu}_\alpha - \sum_n \bar{\psi}_n \gamma^\mu [\partial_\mu - igA^\alpha_\mu t_\alpha]\psi_n - \sum_n m_n \bar{\psi}_n \psi_n \qquad (21)$$

and it is composed almost of the same elements as the QED Lagrangian density in Eq. (16). The new object is the set of eight SU(3) 3×3 matrices $t_\alpha$, numbered by the gluon-color index $\alpha=1,\ldots,8$. They fulfill the SU(3) commutation relations

$$[t_\beta, t_\gamma] = iC^\alpha_{\beta\gamma} t_\alpha, \qquad (22)$$

where $C^\alpha_{\beta\gamma}$ are the SU(3) algebra structure constants [10]. Again, every pair of gluon-color indices implies summation, e.g., over $\alpha$ in Eqs. (21) and (22).

Dirac four-spinors $\psi_n$ correspond to quark fields. Compared to the electron four-spinors $\psi_e$ discussed in Sec. 2.2, they are richer in two aspects. First, each of them appears in three variants, red, blue, and green. These colors are numbered by the quark-color index corresponding to the dimensions 3×3 of the $t_\alpha$ matrices. Traditionally they are not explicitly shown in the Lagrangian density (22), so we should, in fact, think about $\psi_n$ as 12-component spinors. One should not be confused by the fact that there are three colors of quarks, and eight colors of gluons – in fact, here the "visual" representation simply breaks down, and the colors of gluons have nothing to do with red, blue, and green of quarks. In reality, quarks and gluons are numbered by the indices of the corresponding SU(3) representations: three-dimensional spinor representation for quarks, and eight-dimensional vector representation for gluons.

Second, there is not one, but six different quark fields, for $n=1,\ldots,6$. These are called quark flavors, and are usually denoted by names: up, down, charm, strange, top, and bottom, see Fig. 1. For nuclear structure physics, essential rôle is played only by the up and down quarks that are constituents of neutrons and protons. So in most applications of the QCD to nuclear structure, we can limit the QCD Lagrangian density to two flavors only, $n=1, 2$.

Figure 2: Same as in Fig. (1) but for bosons. From `http://www.cpepweb.org/`.

The first term in the QCD Lagrangian density (21) describes the free gluon fields defined by eight four-potentials $A^\alpha_\mu$. One can say that instead of one photon of the QED, that transmits the electromagnetic interaction, we have eight gluons that transmit the strong interaction, see Fig. 2.

The gluon field tensors $F^\alpha_{\mu\nu}$ are defined as

$$F^\alpha_{\mu\nu} = \partial_\mu A^\alpha_\nu - \partial_\nu A^\alpha_\mu + C^\alpha_{\beta\gamma} A^\beta_\mu A^\gamma_\nu. \tag{23}$$

Here comes the really big difference between the QED and QCD, namely, the gluon field tensors contain the third term in Eq. (23). As a result, gluons interact with one another – we can say that they are color-charged, while the photon has no charge. It is easy to see that the third term in Eq. (23) implies the charged gluons. Indeed, the Euler-Lagrange equations corresponding to the free gluon fields do produce the source terms when the Lagrangian density is varied with respect to the gluon fields. (In QED, the free-photon-field Lagrangian depends only on derivatives of the photon fields, and not on the photon fields themselves.)

The last term in (21) describes the six free quarks of masses $m_n$ at rest. This does not mean that isolated quarks can exist in Nature, be accelerated, and have their masses measured by their inertia with respect to acceleration. Each free quark obeys the same Dirac equation as the electron in QED. The Dirac equation is given by the last term and the $\partial_\mu$-term in Eq. (21). Quarks couple to gluons through the color currents,

$$J^\mu_\alpha = -ig \sum_n \bar{\psi}_n \gamma^\mu A^\alpha_\mu t_\alpha \psi_n. \tag{24}$$

We are going to discuss this aspect a few paragraphs below; here we only note that all quarks couple to gluons with the same value of the color charge $g$. We cannot give any numerical value to this parameter, because it depends on energy through the mechanism called renormalization, that we shall not discuss in the present course.

Consequences of the gluon charges are dramatic. Namely, the force carriers now exert the same force as the force they transmit. Moreover, sources of the electromagnetic field depend on currents (20) that involve a small parameter – the electron charge, while gluons constitute sources of the color field without any small parameter. Gluons are not only color-charged, but they also produce very strong color fields.

Let us now consider empty space. In a quantum field theory, we cannot just say that the ground state of the empty space is the state with no quanta – we have to solve the proper field equations, with proper boundary conditions, and determine what is the state of the field. Such a state may or may not contain quanta. In particular, whenever the space has a boundary, the ground state of the field does contain quanta – this fact is called the vacuum polarization effect.

In QED, this is a very well known, and experimentally verified effect. For example, two conducting parallel plates attract each other, even if they are not charged and placed in otherwise empty space (this is called the Casimir effect [11]). One can understand this attraction very easily. Namely, the vacuum fluctuations of the electron field may create in an empty space virtual electron-positron pairs. These charged particles induce virtual polarization charges in the conducting plates (it means virtual photons are created, travel to plates, and reflect from them). Hence, the plates become virtually charged, and attract one another during a short time when the existence of the virtual charges, and virtual photons, is allowed by the Heisenberg principle. All in all, a net attractive force between plates appears.

In QED such effects are extremely weak, because the electron has a small charge and a non-zero rest mass. On the other hand, the QCD gluons are massless, and their strong interaction is not damped by a small parameter. As a result, the QCD vacuum polarization effect is extremely strong, and the empty space is not empty at all – it must contain a soup of spontaneously appearing, interacting, and disappearing gluons. Moreover, in the soup there also must be pairs of virtual quark-antiquark pairs that are also color-charged, and emit and absorb more virtual gluons. It turns out that the QCD ground state of an "empty" space is an extremely complicated object. At present, we do not have any glimpse of a possibility to find the vacuum wave function analytically. Some ideas of what happens are provided by the QCD lattice calculations, see e.g. Ref. [12], in which the gluon and quark fields are discretized on a four-dimensional lattice of space-time points, and the differential field equations are transformed into finite-difference equations solvable on a computer.

An example of such a result is shown in Fig. 3. It presents a frozen-frame image, however, the solution is obtained in space *and* time, and hence we know what happens at different times. One movie is worth thousands photos, so interested students are invited to visit the WEB site indicated in the Figure caption, to see the animation of the complete result. Only then one can appreciate the complexity of appearing structures, with blobs of color charge constantly appearing, disappearing and moving around. The QCD vacuum really resembles a soup of boiling gluons and quarks.

It is now obvious that one cannot expect other solutions of the QCD fields to be any simpler. In particular, solutions corresponding to isolated quarks simply do not exist. One can say that an isolated quark would create so many gluons around it that the complete wave function had not been normalizable. Solutions for quark-antiquark pairs, and for triples of quarks, do exist (we do exist after all – the triples of quarks are nucleons our bodies are built of), but are even more complicated to obtain, even within the QCD lattice calculations. There is no hope, neither there is any reason, to describe composite objects like mesons or nucleons directly from quarks and gluons. This is especially true when we want to use these composite objects to build the next-generation composite objects like nuclei.

Here we arrive at the leading idea of our physical description of the real world. Namely, a physicist always begins by isolating the most important degrees of freedom to describe a given system at a given energy and/or size scale. These degrees of freedom must be compatible with the ones that govern objects at a finer level of description, and must define the degrees of freedom useful at any coarser level of description. However, it is neither useful, nor sensible, nor fruitful, nor doable to overjump different levels. Why bother to derive the structure of a living cell from the unified QCD and electroweak Lagrangian? There are at least seven levels in between: nucleons are built of quarks, nuclei of nucleons, atoms of nuclei and electrons, molecules of atoms, amino acids of molecules, proteins of amino acids, and we did not arrive at a cell yet. Well, we shall not embark here on the philosophy of science; in the following we concentrate on describing how mesons and nucleons are built of quarks, and nuclei of nucleons.

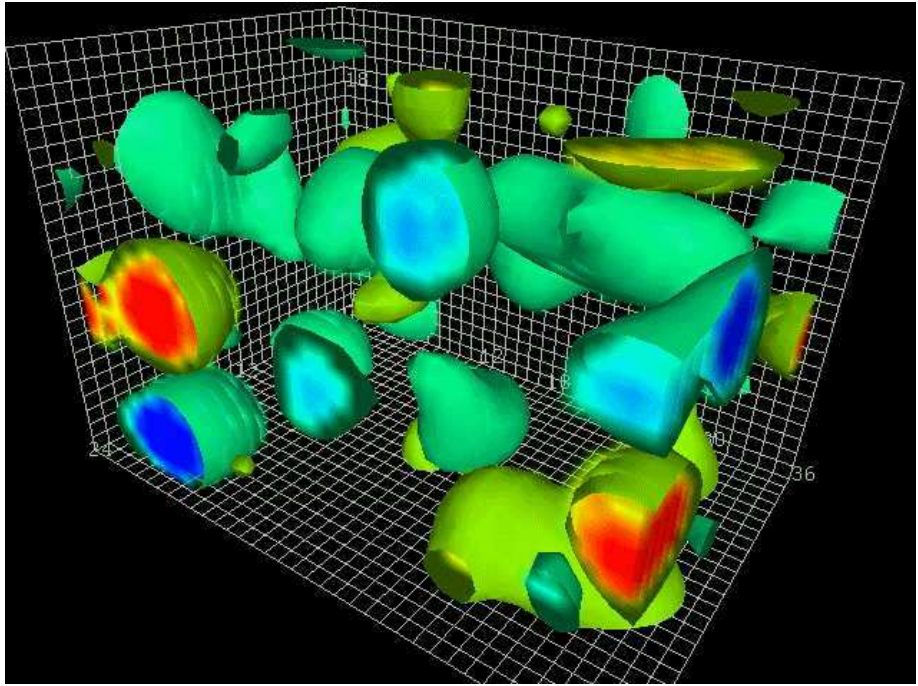

Figure 3: A snap-shot of the space color charge of the QCD vacuum, calculated on a space-time lattice. From `http://hermes.physics.adelaide.edu.au/theory/staff/leinweber/VisualQCD/QCDvacuum/welcome.html`.

## 2.4 Chiral Symmetry and the Isospin

We now proceed with the program outlined at the end of the previous section, namely, knowing from experiment that mesons exist we begin by introducing the relevant degrees of freedom. We also know that meson is a complicated solution of the QCD quark and gluon fields that involve a real quark-antiquark pair. However, without ever being able to find this solution, let us try to identify basic features of the meson that result from the underlying QCD structure.

Let us concentrate on a small piece of the QCD Lagrangian density (21), i.e., on the up and down quark components of the middle term, i.e.,

$$\mathcal{L}_\chi = -\bar{u}\gamma^\mu D_\mu u - \bar{d}\gamma^\mu D_\mu d = -\bar{q}\gamma^\mu D_\mu q. \qquad (25)$$

The gluon fields and the color SU(3) matrices are not essential now, so we have hidden all that in the SU(3) covariant derivative: $D_\mu = \partial_\mu - igA_\mu^\alpha t_\alpha$. On the other hand, we have explicitly indicated the up and down quark fields, $u$ and $d$, and moreover, we have combined both fields into the quark iso-spinor,

$$q = \begin{pmatrix} u \\ d \end{pmatrix}. \qquad (26)$$

To be specific, $q$ contains 24 components, i.e., two quarks, each in three colors, and each built as a four-component Dirac spinor. However, the Dirac and color structure is again not essential, so in the present section we may think about $q$ as two-component spinor. For a moment we have also disregarded the quark mass terms – we reinsert them slightly below.

What is essential now are the symmetry properties of $\mathcal{L}_\chi$. This piece of the Lagrangian density looks like a scalar in the two-component field $q$, i.e., it is manifestly invariant with respect to unitary mixing of up and down quarks. We formalize this observation by introducing the isospin Pauli matrices,

$$\tau_1 = \begin{pmatrix} 0 & 1 \\ 1 & 0 \end{pmatrix}, \quad \tau_2 = \begin{pmatrix} 0 & -i \\ i & 0 \end{pmatrix}, \quad \tau_3 = \begin{pmatrix} 1 & 0 \\ 0 & -1 \end{pmatrix}, \tag{27}$$

and we introduce the unitary mixing of up and down quarks in the language of rotations in the abstract isospin space. And yes, this is exactly the same iso-space that we know very well from nuclear structure physics, where the upper and lower components are the neutron and proton. We come back to that later.

What is slightly less obvious, but in fact trivial to anybody acquainted with the relativistic Lorentz group, is the fact that $\mathcal{L}_\chi$ is also invariant with respect to multiplying the quark fields by the $\gamma_5$ Dirac matrix shown in Eq. (18). This property results immediately from the commutation properties of the $\gamma$ matrices (remember that $\bar{q}=q^+\gamma_0$). So in fact, we have altogether six symmetry generators of $\mathcal{L}_\chi$, namely,

$$\vec{t} = \tfrac{1}{2}\vec{\tau} \quad \text{and} \quad \vec{x} = \gamma_5 \vec{t}, \tag{28}$$

where the arrows denote vectors in the iso-space.

It is now easy to identify the symmetry group of $\mathcal{L}_\chi$. We introduce the left-handed $\vec{t}_L$ and right-handed $\vec{t}_R$ generators,

$$\vec{t}_L = \tfrac{1}{2}(1+\gamma_5)\vec{t} = \tfrac{1}{2}(\vec{t}+\vec{x}) \quad \text{and} \quad \vec{t}_R = \tfrac{1}{2}(1-\gamma_5)\vec{t} = \tfrac{1}{2}(\vec{t}-\vec{x}). \tag{29}$$

Since $(\gamma_5)^2=1$, they fulfill the following commutation relations:

$$[t_{Li}, t_{Lj}] = i\epsilon_{ijk}t_{Lk}, \quad [t_{Ri}, t_{Rj}] = i\epsilon_{ijk}t_{Rk}, \quad [t_{Li}, t_{Rj}] = 0, \tag{30}$$

i.e., $\vec{t}_L$ generates the SU(2) group, $\vec{t}_R$ generates another SU(2) group, and since they commute with one another, the complete symmetry group is SU(2)×SU(2). We call this group chiral.

This result is quite embarrassing, because it is in a flagrant disagreement with experiment. On the one hand, we know very well that particles appear in iso-multiplets. For example, there are two nucleons, a neutron and a proton, that can be considered as upper and lower components of an iso-spinor, and there are three pions, $\pi_+$, $\pi_0$, and $\pi_-$, that can be grouped into an iso-vector. So there is no doubt that there is an isospin SU(2) symmetry in Nature, but, on the other hand, what about the second SU(2) group? In the Lorentz group, the $\gamma_5$ Dirac matrix changes the parity of the field, so if $\gamma_5$ was really a symmetry then particles should appear in pairs of species having opposite parities. This is not so in our world. Nucleons have positive intrinsic parity, and their negative-parity brothers or sisters are nowhere to be seen. Parity of pions is negative, and again, the positive-parity mirror particles do not exist any near the same mass.

So the Nature tells us that the SU(2)×SU(2) symmetry of the QCD Lagrangian must be dynamically broken. It means that the Lagrangian has this symmetry, while the physical solutions do not. We already learned that these physical solutions are very complicated, and we are unable to find them and check what are their symmetries. But we do not really need that – experiment tells us that chiral symmetry must be broken, and hence, we can built theories that incorporate this feature on a higher level of description.

Before we construct a model in which the dynamical symmetry breaking mechanism is explicitly built in (and before we show explicitly what such a symmetry breaking really is), let us first reinsert the quark-mass terms into the discussed piece of the Lagrangian:

$$\mathcal{L}'_\chi = -\bar{u}\gamma^\mu D_\mu u - \bar{d}\gamma^\mu D_\mu d - m_u \bar{u}u - m_d \bar{d}d. \tag{31}$$

By a simple calculation we can now easily show that neither of the two mass terms, nor any linear combination thereof, are invariant with respect to the chiral group SU(2)×SU(2). For certain, had the quark masses been equal, the two combined mass terms would have constituted an isoscalar (an invariant with respect to the isospin group), but even then they would not be chiral scalars (invariants with respect to the chiral group). So the non-zero quark masses break the chiral symmetry. What are the values of these masses has to be taken from the experiment, and indeed, neither the up and down quark masses are zero, nor they are equal to one another, see Fig. 1. The chiral symmetry is therefore broken in two ways: (i) explicitly, by the presence of a symmetry breaking term in the Lagrangian, and (ii) dynamically, as discussed above. Without going into details, we just mention that the non-zero mass of the $\pi$ mesons results from the non-zero quark masses, see Ref. [4], chap. 19.3. For more quark flavors taken into account, the dimensionality of the chiral group increases, i.e., when three quarks $u$, $d$, and $s$ are considered the chiral group is SU(3)×SU(3).

That is about this far that we can move forward by using the QCD quark Lagrangian. We have identified basic symmetry properties of the QCD solutions, and now we have to go to the next level of description, namely, consider composite objects built of quarks. This way of proceeding is called the effective field theory (EFT). We do not build fields of composite objects from the lower-level fields. Instead, we consider the composite objects to be elementary, and we guess their properties from symmetry considerations of the lower-level fields; otherwise, it would have been too difficult a task. Before we arrive at sufficiently high energies, or small distances, at which the internal structure of composite objects becomes apparent, we can safely live without knowing exactly how the composite objects are constructed.

## 2.5 Dynamical (Chiral) Symmetry Breaking

The present subsection is located within the section on quantum fields, but in fact, we tell here a much more general story. Dynamical (or sometimes called – spontaneous) symmetry breaking is a leading theme of a multitude of quantum effects. The very simple model we consider here is a perfect illustration of what is meant by the dynamical symmetry breaking, and moreover it explicitly illustrates the breaking of the chiral symmetry.

### 2.5.1 Non-Linear $\sigma$ Model

The non-linear $\sigma$ model [13, 14] is built to describe pseudoscalar mesons of which we know that: 1° they exist, 2° their scalar partners don't, and 3° they obey the chiral symmetry of SU(2)×SU(2). The first two facts are experimental ones, and the third one comes from the lower level (quark) theory.

The SU(2)×SU(2) group is isomorphic to the O(4) group – the orthogonal group in four dimensions [10]. Therefore, the meson fields in question can be described by four real fields $\phi_n$, $n$=1,2,3,4, and all we need is a model for the Lagrangian density. The non-linear $\sigma$ model makes the following postulate:

$$\mathcal{L}_\sigma = -\tfrac{1}{2}\partial_\mu \phi_n \partial^\mu \phi_n - \tfrac{1}{2}\mathcal{M}^2 \phi_n \phi_n - \tfrac{1}{4}g(\phi_n \phi_n)^2, \tag{32}$$

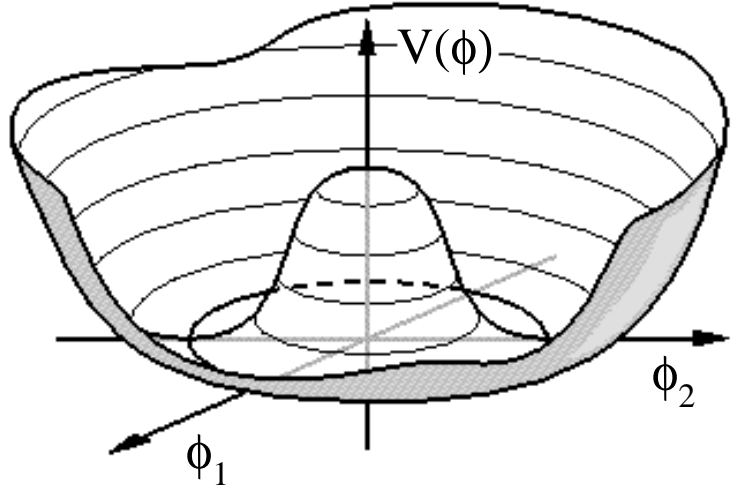

Figure 4: Shape of the "Mexican hat" potential in two dimensions. (Picture courtesy: E.P.S. Shellard, DAMTP, Cambridge.) From http://www.geocities.com/CapeCanaveral/2123/breaking.htm.

where all pairs of indices imply summations. Since only lengths of vectors in the four-dimensional O(4) space appear in the Lagrangian density, it is explicitly invariant with respect to the chiral group.

The potential energy depends only on the radial variable $\sigma^2 = \phi_n \phi_n$, i.e.,

$$V(\phi_n) = V(\sigma) = \tfrac{1}{2}\mathcal{M}^2 \sigma^2 + \tfrac{1}{4} g \sigma^4, \tag{33}$$

but it does not depend on the orientation of $\phi_n$ in the four-dimensional space. For $g>0$ and $\mathcal{M}^2<0$ this potential, as function of $\sigma$, has a maximum at $\sigma=0$, and a minimum at

$$\sigma_0 = |\mathcal{M}|/\sqrt{g}. \tag{34}$$

However, as a function of all the four components $\phi_n$ it is flat in all directions perpendicular to the radial versor $\phi_n/\sigma$. In two dimensions, such a potential is called the "Mexican hat", see Fig. 4.

Let us now consider the classical ground state corresponding to Lagrangian density (32). The lowest energy corresponds to particles at rest, $\partial^\mu \phi_n = 0$, and resting at a lowest point of the potential energy (34). Now we have a problem – which one of the lowest points to choose, because any one such that $\bar{\phi}_n \bar{\phi}_n = \sigma_0^2$ is as good as any other one. However, the classical fields $\phi_n$ at space-time point $x_\mu$ must have definite values, i.e., they *spontaneously* pick one of the solutions $\bar{\phi}_n$ out of the infinitely-many existing ones. Once one of the solutions is picked, the O(4) symmetry is broken, because the ground-state field is not any more invariant with respect to all O(4) transformations. Using the graphical representation of the "Mexican hat", Fig. 4, one can say that the system rolls down from the top of the hat, and picks one of the points within the brim.

It is now clear that fields $\phi_n$ do not constitute the best variables to look at the problem, because the physics in the radial and transversal directions is different. Before proceeding any further, let us introduce variables $\sigma$ and $\vec{z}$ that separately describe these two directions, namely,

$$\vec{\phi} = \frac{2\vec{z}}{1+\vec{z}^2}\sigma \quad \text{for } n=1,2,3, \quad \text{and} \quad \phi_4 = \frac{1-\vec{z}^2}{1+\vec{z}^2}\sigma. \tag{35}$$

Inserting expressions (35) into (32) we obtain the Lagrangian density expressed by the new fields $\vec{z}$ and $\sigma$,

$$\mathcal{L}_\sigma = -2\sigma^2 \frac{\partial_\mu \vec{z} \circ \partial^\mu \vec{z}}{(1+\vec{z}^{\,2})^2} - \tfrac{1}{2}\partial_\mu\sigma\partial^\mu\sigma - \tfrac{1}{2}\mathcal{M}^2\sigma^2 - \tfrac{1}{4}g\sigma^4, \qquad (36)$$

where "$\circ$" distinguishes the scalar product in the iso-space from the scalar product in usual space, which is denoted by "$\cdot$". Apart from the multiplicative factor $\sigma^2$ in front of the first term, the Lagrangian density is now separated into two parts that depend on different variables. Stiffness in the $\sigma$ direction of the potential energy (33), calculated at the minimum $\sigma_0$, equals $m^2 = d^2V/d\sigma^2 = -2\mathcal{M}^2 > 0$, and for large $|\mathcal{M}|$ is very large. Then, the field $\sigma$ is confined to values very close to $\sigma_0$, and we can replace the pre-factor of the first term in Eq. (36) by $\sigma_0$. Within this approximations, fields $\sigma$ and $\vec{z}$ become independent from one another, and can be treated separately.

We disregard now the part of the Lagrangian density depending on $\sigma$. Indeed, the initial potential (33) has been postulated without any deep reason, and a detailed form of it is, in fact, totally unknown – it comes from the quark level that we did not at all solved. Any potential that confines the field $\sigma$ to values close to $\sigma_0$ is good enough. This field must remain in its ground state, because any excitations of it would bring too much energy into a meson, and again, meson's internal structure remains unresolved.

### 2.5.2 Pion-Pion Lagrangian

The remaining fields $\vec{z}$ can be identified with the $\pi$ mesons forming the pseudoscalar isovector multiplet $\vec{\pi} = (\pi_+, \pi_0, \pi_-)$,

$$\vec{\pi} = F_\pi \vec{z}, \qquad (37)$$

where $F_\pi = 2\sigma_0$ is a normalization constant. The pion-pion Lagrangian density then equals

$$\mathcal{L}_{\pi\pi} = -2\sigma_0^2 \frac{\partial_\mu \vec{z} \circ \partial^\mu \vec{z}}{(1+\vec{z}^{\,2})^2} = -\frac{1}{2}\frac{\partial_\mu \vec{\pi} \circ \partial^\mu \vec{\pi}}{(1+\vec{\pi}^{\,2}/F_\pi^2)^2} = -\frac{F_\pi^2}{2}\vec{D}_\mu \circ \vec{D}^\mu, \qquad (38)$$

where we have defined the O(4) covariant derivative

$$\vec{D}_\mu = \frac{\partial_\mu \vec{z}}{1+\vec{z}^{\,2}}. \qquad (39)$$

First of all we notice that Lagrangian density (38) contains only one isovector multiplet of mesons – the parity-inversed chiral partners have disappeared. This is good. The mechanism of the chiral symmetry breaking explains this experimental fact very well. In reality, the chiral partners still exist, but they have been hidden in the $\sigma$ field and pushed up to high excitation energies. They can only be revealed by exciting an (unknown) internal structure of the meson.

Second, Lagrangian density (38) contains no mass term (term proportional to $\vec{z}^{\,2}$), so the pions we have obtained are massless. This is no accident, but a demonstration of a very general fact that for dynamically broken symmetry there must exist a massless boson. This fact is called the Goldstone theorem [15], and the particle is called the Goldstone boson. It sounds very sophisticated, but in fact it is a very simple observation. Even in classical mechanics, if a particle is put into the "Mexican hat" potential and treated within the small-vibration approximation, one immediately obtains a zero-frequency mode that corresponds to uniform motion around the hat. The Goldstone boson is just that.

Third, we have derived the particular dependence of the pion-pion Lagrangian (38) on the derivatives of the pion field. Every such derivative must be combined with the particular denominator to form the covariant derivative $\vec{D}_\mu$ (39). This guarantees the proper transformation properties of the pion field with respect to the chiral group. When we later proceed with constructing other Lagrangian densities of composite particles, we shall use such a dependence on the pion fields.

Experimental masses of pions are not equal to zero, so the obtained pion-pion Lagrangian density is too simplistic. However, we can now recall that the quark mass terms do break the chiral symmetry *explicitly* (see Sec. 2.4). This corresponds to a slight tilt of the "Mexican hat" to one side. (To which side, is perfectly well defined by the O(4) structure of the quark mass terms in Eq. (31) – but we shall not discuss that.) Such a tilt creates a small curvature of the potential along the valley within the hat's brim, and this curvature gives the pion-mass term $-\frac{1}{2}m_\pi^2 \vec{\pi}^2$ in the pion-pion Lagrangian density. So the non-zero quark masses result in a non-zero pion mass. By the way, the difference in masses of neutral and charged pions results from a coupling to virtual photons – its origin is therefore in the QED, and not in the QCD.

It is amazing how much can be deduced from considerations based on the idea of the dynamical symmetry breaking. Considering the complication of the problem, that is unavoidable on the quark-gluon level, we have reached important results at a very low cost. This happens again and again in almost every branch of physics of the micro-world. Dynamical breaking of the local gauge symmetry gives masses to the electroweak bosons $Z^0$ and $W^\pm$, and leaves the photon massless. Dynamical breaking of the rotational symmetry in nuclei creates the collective moment of inertia and rotational bands. Dynamical breaking of the particle-number symmetry gives superconducting condensates in nuclei and in crystals. Dynamical breaking of the parity symmetry in nuclei and molecules gives collective partner bands of opposite parities. Dynamical breaking of the chiral symmetry (in a different sense, pertaining to the time-reversal symmetry) has been suggested to explain pairs of nuclear rotational bands having the same parity. The story just does not end. Dynamical symmetry breaking rules the world.

### 2.5.3 Nucleon-Pion Lagrangian

We are now ready to consider another set of composite particles, the nucleons. We know that there are two nucleons in Nature, of almost equal mass, the neutron and the proton, so they can be combined into the iso-spinor

$$N = \begin{pmatrix} p \\ n \end{pmatrix}, \qquad (40)$$

where $p$ and $n$ are the Dirac four-spinors of spin 1/2 particles. We have already attributed the isospin projections to quarks, Eq. (26), by placing within the quark iso-spinor the quark up up and the quark down down (sounds logical?). Since the proton is made of the (*uud*) quarks, and the neutron of the (*udd*) quarks, their isospin projections are therefore determined as in Eq. (40). In nuclear structure physics one usually uses the opposite convention, attributing the isospin projection $t_3 = +\frac{1}{2}$ to a neutron, in order to make most nuclei to have positive total isospin projections $T_3 > 0$. All this is a matter of convention; one could as well put the quark up down and the quark down up – the physics does not depend on that.

Anyhow, the nucleons contain not only the three (valence) quarks, but also plenty of gluons, and plenty of virtual quark pairs, and we are unable to find what exactly this state is. Therefore, here we follow the general strategy of attributing elementary fields to composite particles. Before we arrive at sufficiently high energies, or small distances, at which the internal structure of composite objects becomes apparent, we can safely live without knowing exactly how the composite objects are constructed.

As usual, having defined elementary fields of particles that we want to describe, we also have to postulate the corresponding Lagrangian density. And as usual, we do that by writing a local function of fields that is invariant with respect to all conserved symmetries. When we have the nucleon and pion fields at our disposal, and we want to construct the Lorentz and chiral invariant Lagrangian density, the answer is:

$$\mathcal{L}_{N\pi} = -\bar{N}\left(\gamma^\mu \partial_\mu + g_\phi \left[\phi_4 + 2i\gamma_5 \vec{t} \circ \vec{\phi}\right]\right) N. \tag{41}$$

If you are not tired of this game of guessing the right Lagrangian densities, you may wonder why the meson fields (within the square brackets) appear in this particular form. To really see this, we have to recall more detailed properties of the chiral group SU(2)×SU(2). Its generators $\vec{t}$ and $\vec{x}$ in the spinor representation are given by Eq. (28), however, when more than one quark is present, we have to use the analogous generators $\vec{T}$ and $\vec{X}$ that are sums of $\vec{t}$'s and $\vec{x}$'s for all quarks. In particular, the meson fields $\phi_n$ belong to the vector representation of SU(2)×SU(2). Then, according to identification (35) and (37), the first three components $\vec{\phi}$ form the isovector pion field, and the fourth component $\phi_4$ is an isoscalar. This fixes the transformation properties of $\phi_n$ with respect to the iso-rotations, given by infinitesimal transformation $-i\vec{\theta}\circ\vec{T}$. Since these rotations have identical form as the real rotations in our three-dimensional space, we do not show them explicitly. On the other hand, the transformation properties of $\phi_n$ with respect to the chiral rotations, given by infinitesimal transformation $-i\vec{\epsilon}\circ\vec{X}$, are

$$\vec{\phi} \longrightarrow \vec{\phi} + \vec{\epsilon}\phi_4 \quad \text{and} \quad \phi_4 \longrightarrow \phi_4 - \vec{\epsilon}\circ\vec{\phi}. \tag{42}$$

There is no magic in this expression – one only has to properly identify generators of the O(4) group with generators $\vec{T}$ and $\vec{X}$. This is unique, once we fix which components (1,2,3 in our case) transform under the action of $\vec{T}$. Under the chiral rotation about the same angle $\vec{\epsilon}$, the nucleon fields transform by infinitesimal transformation $-i\vec{\epsilon}\circ\vec{x}$, within the spinor representation of Eq. (28), i.e.,

$$N \longrightarrow N - i\gamma_5 \vec{\epsilon}\circ\vec{t}N \quad \text{and} \quad \bar{N} \longrightarrow \bar{N} - \bar{N}i\gamma_5\vec{\epsilon}\circ\vec{t}. \tag{43}$$

It is now a matter of a simple algebra to verify that Lagrangian density (41) remains invariant under chiral rotations of fields (42) and (43). Note that the first term in Eq. (41) is separately chiral invariant, so we could multiply the second term by an arbitrary constant $g_\phi$.

We can now proceed with the transformation to better variables $\sigma$ and $\vec{z}$, given by Eq. (35), which gives,

$$\mathcal{L}_{N\pi} = -\overline{\widetilde{N}}\left(\gamma^\mu\partial_\mu + g_\phi\sigma + 2i\vec{t}\circ(\vec{z}\varowedge\gamma^\mu\vec{D}_\mu) + 2ig_A\gamma_5\vec{t}\circ\gamma^\mu\vec{D}_\mu\right)\widetilde{N}, \tag{44}$$

where $\varowedge$ denotes vector product in the iso-space. Covariant derivatives of pion fields $\vec{D}_\mu$ are defined as in Eq. (39), and the chiral-rotated nucleon field $\widetilde{N}$ is defined as

$$\widetilde{N} = \frac{(1 + 2i\gamma_5\vec{t}\circ\vec{z})N}{\sqrt{1+\vec{z}^2}}. \tag{45}$$

There are several fantastic results obtained here. First of all, the nucleon mass term $-m_N\overline{\widetilde{N}}\widetilde{N}$ appears out of nowhere, and the nucleon mass,

$$m_N = g_\phi\sigma_0, \tag{46}$$

is given by the chiral-symmetry-breaking value $\sigma_0$ of the $\sigma$ field. In principle, we could begin by including the nucleon mass term already in the initial Lagrangian density (41). This is not necessary – the nucleon mass results from the same chiral-symmetry-breaking mechanism that pushes scalar mesons up to high energies. Second, the third term in Eq. (44) gives the coupling of nucleons to mesons, and in the potential approximation it yields the long-distance, low-energy tail of the nucleon-nucleon interaction, i.e., the one-pion-exchange (OPE) Yukawa potential [16]. Derivation of this potential from Lagrangian density (44) requires some fluency in the methods of quantum field theory, so we do not reproduce it here. Suffice to say, that the OPE potential appears as naturally from exchanging pions, as the Coulomb potential appears from exchanging photons via the electron-photon coupling term in Eq. (16). Last but not least, the last term in Eq. (44) gives the axial-vector current that defines the weak coupling of nucleons to electrons and neutrinos. From where phenomena like the $\beta$ decay can be derived. [This term is an independent chiral invariant, so again we could put a separate coupling constant there; experiment gives $g_A$=1.257(3).]

# 3 FEW-NUCLEON SYSTEMS

In the previous section we have obtained Lagrangian densities that describe composite particles like pions (38) or nucleons (44). These particles are built of the $u$ and $d$ valence quarks as well as of virtual gluons and quark-antiquark pairs. In fact, we can say that the virtual constituents provide for binding of the valence constituents, and most of the rest mass of composite particles comes from the binding by virtual constituents. Indeed, the rest masses of pions, $m_{\pi^0} \simeq 135$ MeV and $m_{\pi^\pm} \simeq 140$ MeV, and nucleons, $m_p \simeq 938$ MeV and $m_n \simeq 940$ MeV, are much, much larger than those of quarks, $m_u \simeq 3$ MeV and $m_d \simeq 6$ MeV. Moreover, the famous confinement effect prevents the valence quarks from being separated one from another, unless a real quark-antiquark pair is created from the vacuum, and two separate composite particles appear.

It is amazing that strong interactions only yield a strong binding for objects that cannot be broken apart at all. Once these composite particles are built, strong interactions become almost completely saturated, and what remains of them, when looked upon from the outside of composite particles, is in fact a relatively weak force.

Let us illustrate this weakness of the strong force by several examples. First of all, the only bound binary system of nucleons, i.e., the deuteron $n$-$p$, has the binding energy of only $B_D$=2.224575(9) MeV. This is really a small number as compared to, e.g., either the nucleon rest masses, or the QCD coupling constant. The deuteron is barely bound, and moreover, neither the di-neutron ($n$-$n$), nor the di-proton ($p$-$p$) is a bound object. The $n$-$p$ scattering amplitude has a pole (corresponding to the deuteron bound state) at the relative momentum of only $k=i\sqrt{m_N B_D} \simeq 45i$ MeV (the pole appears on the imaginary axis). The corresponding scattering length is fairly large, $a$=5.424(3) fm, and certainly much larger than the size of the deuteron, $R_D$=1.953(3) fm. The range of the nucleon-nucleon (NN) interaction, as given by the OPE potential, corresponds to the inverse of the pion mass $1/m_\pi$, and equals to about 1.4 fm [note that in the units $\hbar = c = 1$ we have $1\,\mathrm{fm} \simeq (197\,\mathrm{MeV})^{-1}$].

The above scattering characteristics pertain to the $^3S_1$ channel, i.e., to a scattering with the total spin of $S$=1, the total orbital angular momentum of $L$=0, the total angular momentum of $J$=1, and the total isospin of $T$=0. In the $^1S_0$ channel ($S$=0, $L$=0, $J$=0, and $T$=1) the deuteron is unbound, the scattering amplitude has a pole at $k \simeq -8i$ MeV (on the negative imaginary axis – corresponding to the so-called virtual, or quasibound state), and the scattering length is very large negative, $a$=−23.749(8) fm. The $n$-$n$ and $p$-$p$ scattering lengths (in the $^1S_0$ channel) are also very large negative, $a$=−18.5(4) fm and $a$=−7.806(3) fm, respectively. Finally, the multi-nucleon bound objects (i.e., the atomic nuclei) are also very weakly bound, with the binding energy per nucleon of only $B/A \simeq 8$ MeV.

These weak bindings have very important consequences for the physics of nuclear systems, namely, in these systems, the average kinetic energies are large positive, and the average potential (interaction) energies are large negative. The resulting total energies are therefore much smaller than either the kinetic or the potential component. As a result, one can neither treat the interaction as a small perturbation on top of the (almost) free motion of constituents, nor treat the relative kinetic energy as a small perturbation on top of a tightly bound, frozen system.

## 3.1 Nucleon-Nucleon Interaction

Let us discuss in some more detail the interaction between nucleons. In the past there has been a tremendous experimental effort devoted to scattering protons on protons and neutrons on protons. Since the neutron target is not available, the neutron-neutron scattering was inferred mostly from the scattering of protons on deuterons. All this effort lead to a large database of cross-sections and phase shifts that provide the most extensive information on the binary interactions on nucleons. There have also been numerous attempts to model the interaction between nucleons by different kinds of potentials. Here we limit the discussion to the Argonne $v_{18}$ potential [17], and refer the reader to this paper for references to other existing approaches.

The Argonne $v_{18}$ NN interaction consists of the electromagnetic $V^{\text{EM}}$, one-pion-exchange $V^\pi(NN)$, and intermediate and short-range phenomenological $V^{\text{R}}(NN)$ parts, i.e.,

$$V(NN) = V^{\text{EM}}(NN) + V^\pi(NN) + V^{\text{R}}(NN). \tag{47}$$

The electromagnetic part contains not only the standard Coulomb interaction between protons, but also various other terms like the two-photon Coulomb terms, vacuum polarization terms, and magnetic-moment interactions. The OPE potential results directly from the nucleon-pion Lagrangian discussed in Sec. 2.5.3, and has the following explicit form (here shown for the $p$-$p$ interaction):

$$V^\pi(pp) = f_{pp}^2 \tfrac{1}{3} m_\pi \left[ Y(r)\boldsymbol{\sigma}_i \cdot \boldsymbol{\sigma}_j + T(r)S_{ij} \right], \tag{48}$$

where

$$Y(r) = \frac{e^{-m_\pi r}}{r}\left(1 - e^{-cr^2}\right), \tag{49}$$

$$T(r) = \left(1 + \frac{3}{m_\pi r} + \frac{3}{(m_\pi r)^2}\right)\frac{e^{-m_\pi r}}{r}\left(1 - e^{-cr^2}\right)^2, \tag{50}$$

and $S_{ij}=3(\boldsymbol{\sigma}_i \cdot \boldsymbol{r})(\boldsymbol{\sigma}_j \cdot \boldsymbol{r})/r^2 - \boldsymbol{\sigma}_i \cdot \boldsymbol{\sigma}_j$ is the tensor operator which depends on the Pauli matrices of the $i$th and $j$th interacting particles. The standard OPE terms have been supplemented with the cut-off factors $(1 - e^{-cr^2})$ that kill these terms at distances smaller than $r_c=1/\sqrt{c}$, i.e., below $r_c=0.69$ fm for the used value of $c=1.21$ fm$^{-2}$. There, the remaining terms come into play:

$$V^{\text{R}} = V^{\text{c}} + V^{\text{l2}}L^2 + V^{\text{t}}S_{12} + V^{\text{ls}}L \cdot S + V^{\text{ls2}}(L \cdot S)^2, \tag{51}$$

where the i=c, l2, t, ls, and ls2 terms read

$$V^{\text{i}}(r) = I^{\text{i}}T^2(r) + \left[P^{\text{i}} + Q^{\text{i}}m_\pi r + R^{\text{i}}(m_\pi r)^2\right]\left(1 + e^{(r-r_0)/a}\right)^{-1}, \tag{52}$$

and $I^{\text{i}}$, $P^{\text{i}}$, $Q^{\text{i}}$, and $R^{\text{i}}$ are parameters fitted to the scattering data. These terms are cut off at large distances, i.e., above $r_0=0.5$ fm, with the transition region of the width of $a=0.2$ fm.

The Argonne $v_{18}$ potential adopts the point of view that at large distances the NN interaction is governed by the OPE effects, while the short-range part is treated fully phenomenologically.

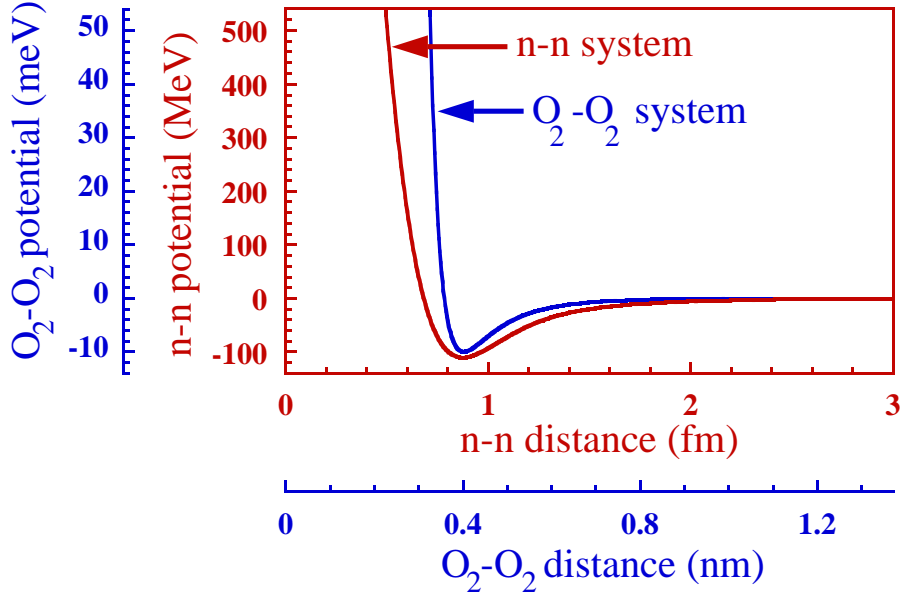

Figure 5: The $^1S_0$-channel $n$-$n$ potential in megaelectronovolts (MeV), as function of the distance in femtometers (fm) (inner axes) compared with the $O_2$-$O_2$ molecular potential in millielectronovolts (meV), as function of the distance in nanometers (nm) (outer axes).

In this respect, there is a perfect analogy between the strong force acting between nucleons, as modelled by Argonne $v_{18}$, and the electromagnetic force acting between neutral non-polar molecules, modelled by the Lennard-Jones potential.

Nucleons are colorless objects, i.e., when looked upon from the outside; no net color charge is visible. The same is true for neutral non-polar molecules that contain equal amounts of positive and negative electromagnetic charges distributed with no net shift, and hence they have no net charge or dipole moment. However, when two molecules approach one another, the charges become polarized, and each molecule acquires a non-zero dipole moment. Then the leading-order interaction energy between molecules equals $V(\boldsymbol{r}) = -2\boldsymbol{E}(\boldsymbol{r}) \cdot \boldsymbol{d}(\boldsymbol{r})$, where $\boldsymbol{E}(\boldsymbol{r})$ is the average electric field felt by one of the molecules when the second one is located at $\boldsymbol{r}$, and $\boldsymbol{d}(\boldsymbol{r})$ is its dipole moment. Assuming that the induced dipole moment $\boldsymbol{d}(\boldsymbol{r})$ depends linearly on the electric field, and knowing that the electric field created by a dipole decreases as $1/r^3$, we obtain immediately that $V(\boldsymbol{r}) \sim -1/r^6$, which gives the well-known Van der Waals potential. At intermediate and small distances, polarization effects become stronger, and higher induced multipole moments begin to be active, however, we can model these effects by a phenomenological term that is equal to the square of the Van der Waals term. Together, one obtains the Lennard-Jones potential,

$$V_{\rm LJ}(r) = 4E_{p,0}\left[\left(\frac{\sigma}{r}\right)^{12} - \left(\frac{\sigma}{r}\right)^{6}\right], \tag{53}$$

where $E_{p,0}$ and $\sigma$ are parameters fitted to data.

In Fig. 5 we show a comparison of the $n$-$n$ Argonne $v_{18}$ potential in the $^1S_0$ channel, with the Lennard-Jones potential between two $O_2$ molecules ($E_{p,0}$=10 meV and $\sigma$=0.358 nm). The Argonne $v_{18}$ potential has been calculated by using the `av18pw.f` FORTRAN subroutine [17], available at http://www.phy.anl.gov/theory/research/av18/av18pot.f. Both potentials are drawn in the same Figure with two abscissas (the lower one for $O_2$-$O_2$, the upper one for $n$-$n$) and two ordinates (the left one for $O_2$-$O_2$, the right one for $n$-$n$). Scales an the abscissas were fixed so as to put the minima of potentials at the same point, and differ by a factor of about $0.5 \times 10^6$, while scales on the ordinates differ by the factor of $10^{10}$.

Despite the tremendous differences in scales, both potentials are qualitatively very similar. Amazingly, it is the electromagnetic molecule-molecule potential that it stiffer at the minimum than the neutron-neutron "strong" potential. In this respect, it is fully justified to put the word "strong" into quotation marks – this potential is not strong at all! Both potentials exhibit a very strong repulsion at short distances – the so-called hard core (the $O_2$-$O_2$ repulsion is stronger!). At large distances, there appears a weak attraction (the $n$-$n$ attraction vanishes more slowly – despite the exponential form of the OPE potential). Neither of the potentials is strong enough to bind the constituents into a composite object.

The analogy between the "strong" NN force and the electromagnetic molecule-molecule force is extremely instructive. First of all, we can demystify the OPE potential in the sense that the exchange of real particles (pions) is, in fact, *not* its essential element. The OPE potential is a remnant of our tool (quantum field theory) that we used to derive it, but on a deeper level it is an effect of the color force between color-polarized composite particles. After all, nobody wants to interpret the dipole-dipole inter-molecular $O_2$-$O_2$ force by an exchange of a "particle". This force can be understood in terms of a more fundamental interaction – the Coulomb force. Second, although the asymptotic, large-distance, leading-order behaviour of both potentials can fairly easily be derived, at intermediate and small distances the interaction becomes very complicated. This is not a reflection of complications on the level of fundamental forces (color or electromagnetic), but a reflection of the complicated polarization effects that take place when composite objects are put close to one another. Moreover, these polarization effects have *per se* quantum character, because the fermionic constituents do not like being put close to one another – the Pauli exclusion principle creates additional polarization and repulsion effects. And third, it is obvious that at small distances there must appear effects that are of a *three-body* character. Namely, when three $O_2$ molecules approach each other (e.g., in liquid oxygen), the basic assumption that they polarize one another only in pairs does not hold. There are certainly polarization effects that depend on explicit positions of the three of them. Similarly, when three nucleons approach each other within the nucleus, their quark-gluon magma becomes polarized in a fairly complicated way, which on the level of potential energy (total-interaction energy) reveals additional terms depending on the three positions simultaneously; this gives the three-body NNN force.

## 3.2 The Deuteron, and more about the Dynamical Symmetry Breaking

Having defined the two-body force that acts between the nucleons, we can relatively easily find the ground-state wave function of the deuteron, and calculate all its properties. In doing so one cannot forget that for $S=1$ states, the tensor terms in the interaction can mix interaction channels, i.e., for any angular momentum $J>0$, states with $L=J\pm1$ are mixed if their parity equals $\pi=-(-1)^J$. These conditions are fulfilled for the $J^\pi=1^+$ deuteron ground state, and hence interactions channels $^3S_1$ and $^3D_1$ contribute to the deuteron ground-state wave function.

The solution corresponding to the Argonne $v_{18}$ interaction is illustrated in Fig. 6, where surfaces of equal density are shown for the $M_J=0$ and 1 magnetic substates of the $J=1$ deuteron ground state. Interested students are invited to visit the WEB site indicated in the Figure caption, to see the animation that shows similar surfaces at other densities. The surfaces are here shown by stripes that allow seeing the other side of the deuteron. The colors are used only to enhance the three-dimensional rendition of the image, and have no other meaning. In particular, the fact that the front piece of the left part in Fig. 6 is red, and the rear piece is blue, does not mean that the neutron is represented in red and the proton in blue, nor that it has been rendered the other way around. In reality, the laboratory-frame wave function has $T$=0, i.e, it is an antisymmetrized combination of products of the neutron and proton wave functions.

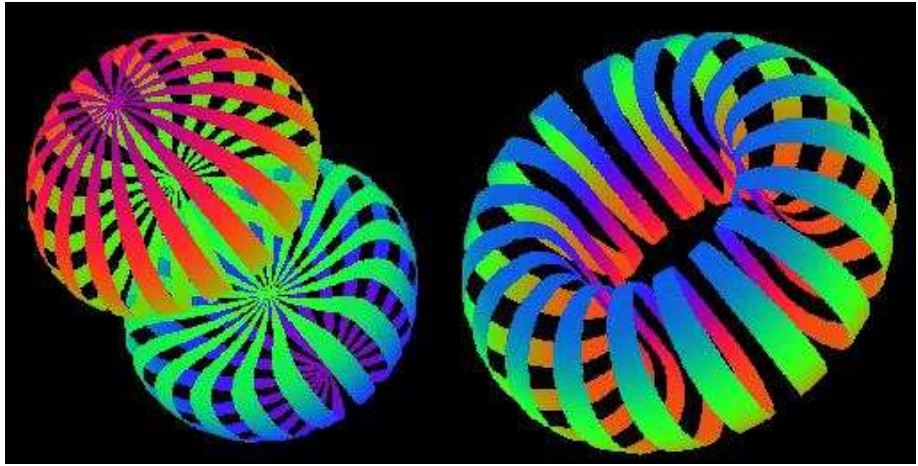

Figure 6: Shapes of the deuteron in the laboratory reference frame. Stripes show surfaces of equal density for the $M_J$=1 (left) and $M_J$=0 (right) magnetic substates of the $J$=1 ground state. From http://www.phy.anl.gov/theory/movie-run.html.

This brings us to a very important point pertaining to the dynamical (or spontaneous) symmetry breaking mechanism discussed already in Sec. 2.5. Suppose that you are confronted with a request: *S'il vous plaît...dessine-moi un deuton!* (see Ref. [18] for an analogous example). Without any deep information about the interaction, you would draw to points (or spheres, if you know something about quantum mechanics), some distance apart, and mark one of them with a $p$ and the other one with an $n$. And this is what the deuteron really looks like in the so-called intrinsic reference frame.

One should not attribute too much importance to the descriptions "laboratory frame" and "intrinsic frame". Below we shall use this names at will, but let us rather treat them as proper names describing two different ways of constructing the wave functions, and not as mathematically sound representations of the same wave function in two different reference frames.

The intrinsic wave function of the deuteron breaks the rotational symmetry, and breaks the isospin symmetry, i.e., a rotation in the real space, and a rotation in the iso-space, gives another wave function. In a more mathematical language, such a wave function does not belong to any single representation of the rotational and isospin symmetry groups. You should not be confused by the fact that the laboratory-frame $J$=1 wave function has three magnetic components (two of them are illustrated in Fig. 6), and hence none of them is strictly invariant with respect to the real-space rotations. However, each magnetic component, when rotated, is equal to some linear combination of all magnetic components, i.e., the $J$=1 state is invariant with respect to rotations in this more general sense – it belongs to one, single representation of the rotation group.

Before discussing the sense of the intrinsic wave functions, let us give two other examples of the symmetry-broken intrinsic wave functions. Imagine the ground-state wave function of the water molecule $H_2O$. We know very well how this molecule looks like – the two hydrogen atoms are connected by chemical bonds to the oxygen atom, and the two lines connecting the H and O nuclei form an angle of about 105°. So the wave function of the water molecule breaks the rotational invariance. However, if we take such an isolated molecule, and wait long enough for all its rotational and vibrational excitations to de-excite by the emission of electromagnetic radiation, the molecule will reach the ground state of $J^\pi=0^+$, i.e., the state which is perfectly invariant with respect to rotations.

There is no contradiction between these two pictures of the molecule. The first one pertains to the wave function in the intrinsic reference frame, and the second one to the wave function in the laboratory reference frame. The intrinsic wave function *is not an exact ground state* of the rotationally invariant Hamiltonian. It is a wave packet, which has a good orientation in space, and a very broad distribution of different angular momenta, corresponding to the ground-state rotational band of the water molecule. On the contrary, the laboratory-frame wave function *is an exact ground state* of the rotationally invariant Hamiltonian, it has a definite value of the angular momentum, $J=0$, and has a completely undefined orientation in space.

As the second example, consider the ground state of the $^{166}$Er nucleus. It is a well-deformed nucleus, having the intrinsic ground-state wave function in the form of a cigar (prolate shape), which breaks the rotational symmetry. At the same time, the laboratory ground state has $J^\pi=0^+$, and is perfectly rotationally invariant. Again, the cigar-shape, intrinsic wave function is a wave packet that is oriented in space and has an undefined angular momentum, while the laboratory wave function is an exact eigenstate having a definite angular momentum.

Now comes a very important question, namely, is there anything else in the phenomenon of the dynamical symmetry breaking apart from the trivial wave-packet formation? The answer is, of course, yes! The point is that some systems can, and some other ones cannot be oriented. The first ones do break the symmetry dynamically, and the second ones do not. It is obvious that the water molecule does it. In other words, its moment of inertia is so huge that the ground-state rotational band is very much compressed (compared to other possible excitations), and all rotational states of this band (all different angular momenta) are very close to one another. The wave packet built of such states is therefore "almost" an eigenstate – at least it has a very long lifetime before it decays to the ground state. Hence, the oriented state of the water molecule is a very good rendition of the exact ground state.

On a different scale, the same is true for the $^{166}$Er nucleus. States of its ground-state rotational band live some nanoseconds, i.e., much longer that any other excitations available in this system. Hence, this nucleus can be oriented, and the corresponding wave packet fairly well represents the ground state. This representation is better or worse depending on which observable we want to look at. For example, if we measure the nuclear root-mean-square radius, the oriented wave function can be used at marvel. The increase of radii of deformed nuclei as compared to their spherical neighbours is a very well established experimental fact. Similarly, lifetimes of the rotational states can be very well approximated by the probability of emitting classical radiation from a rotating charged deformed body.

So we can really say that the ground-state $J^\pi=0^+$ wave function of $^{166}$Er does break, and that of $^{208}$Pb does not dynamically break the rotational symmetry. The latter nucleus does not have any rotational band and thus the oriented wave packet cannot exist. Both $J^\pi=0^+$ ground-state wave functions are perfectly rotationally invariant, while the dynamical symmetry breaking is a notion pertaining to their intrinsic structure.

The utility of the intrinsic wave function does not end at systems that dynamically break the symmetry. Namely, often it is very easy to construct approximated symmetry-broken wave functions, and then use its symmetry-projected component to model the exact symmetry-invariant ground state. The deuteron wave function, with which we have begun this discussion, is a perfect example of such a situation. Namely, the intrinsic-frame image of this nucleus (neutron here and proton there) breaks the isospin symmetry, but the component projected on $T=0$ is a very good representation of the exact wave function. In this case, projection on $T=0$ simply means antisymmetrizing the two components with the neutron and proton positions exchanged. The

$T$=0 projected component serves us well, even if the $T$=1 component ($J$=0) is unbound at all.

Moreover, the intrinsic-frame image of the deuteron explains very well why this particle has apparently so different shapes depending on the value of the magnetic projection $M_J$. The $M_J$=0, torus-like shape, Fig. 6, results simply from projecting the intrinsic wave function on $J$=1 and $M_J$=0, which corresponds to taking a linear superposition of all intrinsic states rotated around the axis perpendicular to the line connecting the neutron and proton in the intrinsic frame. Without such an interpretation, nobody would actually believe that deuteron looks like a torus.

## 3.3 Effective Field Theory

As we have discussed, the Argonne $v_{18}$ interaction uses the OPE potential at large distances, and the phenomenological interaction at intermediate and small distances. One can also follow the standard ideology of the quantum-field theory, and model the second piece by the exchange effects for heavier mesons. Larger meson masses mean shorter distances of the interaction, so we can understand why adding more mesons, and using the corresponding Yukawa interactions, we can parametrize the NN force equally well.

Although this way of proceeding works very well in practice, it creates two conceptual problems. First, one has to include the scalar-isoscalar meson called $\sigma$, which has the quantum numbers of a pair of pions. It fulfills the role of an exchange of the a pair of pions, however, such a meson neither exist in Nature as a free particle, nor its mass, that has to be used in the corresponding Yukawa term, is close to the doubled pion mass. The exchange of such a virtual particle simply corresponds to higher-order effects in the exchange of pions, which is a perfectly legitimate procedure, but it departs from the idea that real, physical particles mediate the NN interaction.

Second, two other heavy mesons have to be included, namely, the vector isovector meson $\rho$ and the vector isoscalar meson $\omega$. They are physical particles, with the rest masses of about 800 MeV, and the corresponding ranges of the Yukawa potentials are very small, of the order of 0.25 fm. These small ranges allow to model the NN interaction at very short distances, but at these distances nucleons really start to touch and overlap. Therefore, it is rather unphysical to think that nucleons can still interact as unchanged objects, by exchanging physical particles. Within the image of the strong color-polarization taking place at such a small distances, one would rather think that the internal quark-gluon structure of nucleons becomes strongly affected, which creates strong repulsion effects, predominantly through the Pauli blocking of overlapping quark states.

At present, we are probably not at all able to tell what happens with the nucleons when they are put so near to one another. However, we do not really need such a complete knowledge when describing low-energy NN scattering and structure of nuclei. All what we need is some kind of parameterization of the short-range, high-energy effects when we look at their influence on the long-range, low-energy observables. Such separation of scales is at the heart of the effective field theory (EFT).

One can apply similar ideas to almost all physical systems, where our knowledge of the detailed structure is neither possible nor useful. The simplest example is the effect of the electromagnetic charge and current distributions inside a small object, when we shine at it an electromagnetic wave of a much longer length (the long-wave-length limit). It is well known that all what we then need, are a few numbers – low-multiplicity electric and magnetic moments. Of course, the best would be to be able to calculate these moments from the exact charge and current distributions, but once we know these numbers, we know everything. On the other hand, if the internal structure is not known, we can fit these numbers to the measured long-wave scattering, and thus obtain the complete information needed to describe such a scattering process.

Examples of other such situations are plenty in physics. Interested students are invited to go through very good introductory lecture notes by Lepage [19], where nice instructive examples are presented within the framework of ordinary quantum mechanics. In particular, it is shown how a short-range perturbation of the ordinary Coulomb potential influences the hydrogen atomic wave functions, and how such a perturbation (no matter its physical origin) can be parametrized by a zero-range, delta-like potential.

Here we only discuss two applications of the EFT, which pertain to the pion and nucleon systems. First, let us consider the pion-pion Lagrangian density (38). When we use the methods of the quantum field theory to derive the $\pi$-$\pi$ scattering amplitude, it turns out that details of the experimental results are not well reproduced. This suggests that even during a low-energy scattering process of composite particles, the internal, short-range, high-energy structure does become visible. The question is whether one can modify the Lagrangian density in such a way that the internal quark-gluon degrees of freedom do not explicitly appear, and yet their influence on the $\pi$-$\pi$ scattering amplitude is taken into account. The EFT prescription suggests that one should add to the Lagrangian higher-order terms that depend on the pion field and conserve all symmetries of the theory (Lorentz and chiral invariance in this case). We than obtain the effective Lagrangian density,

$$\begin{aligned}\mathcal{L}_{\pi\pi}^{\text{eff}} = & -\tfrac{1}{2}F_\pi^2 \vec{D}_\mu \circ \vec{D}^\mu - \tfrac{1}{2}m_\pi^2 \frac{\vec{\pi}^2}{1+\vec{\pi}^2/F_\pi^2} \\ & -\tfrac{1}{4}c_4 \left(\vec{D}_\mu \circ \vec{D}^\mu\right)^2 - \tfrac{1}{4}c_4' \left(\vec{D}_\mu \circ \vec{D}_\nu\right)\left(\vec{D}^\mu \circ \vec{D}^\nu\right) + \ldots, \end{aligned} \quad (54)$$

built from the covariant derivatives of the pion field (39), which ensures the chiral invariance, and with all the Lorentz indices summed up in pairs, which ensures the Lorentz invariance. Up to these rules, there are two quartic terms possible, and the series could be, in principle, continued to even higher orders. However, by adjusting free parameters $c_4$ and $c_4'$ we are now able to properly describe the experimental $\pi$-$\pi$ scattering data. Note that the quartic terms in the local Lagrangian density can be interpreted as zero-range contact (point-like) interactions. In Eq. (54) we have also included the pion mass term $m_\pi$, which explicitly (but weakly) breaks the chiral invariance.

The second example concerns the nucleon-pion Lagrangian density (44) that can be transformed into an effective Lagrangian by adding terms which are quartic in the nucleon fields,

$$\begin{aligned}\mathcal{L}_{N\pi}^{\text{eff}} = & -\overline{\widetilde{N}}\left(\gamma^\mu \mathcal{D}_\mu + m_N + 2ig_A\gamma_5 \vec{t} \circ \gamma^\mu \vec{D}_\mu\right)\widetilde{N} \\ & - c_{2\alpha\beta}\left(\overline{\widetilde{N}}\Gamma_\alpha \widetilde{N}\right)\left(\overline{\widetilde{N}}\Gamma_\beta \widetilde{N}\right) + \ldots, \end{aligned} \quad (55)$$

where we have combined two terms of Eq. (44) into the covariant derivative of the nucleon field,

$$\mathcal{D}_\mu = \partial_\mu + 2i\vec{t}\circ(\vec{z}\boxtimes\gamma^\mu\vec{D}_\mu). \quad (56)$$

Symbols $\Gamma_\alpha$ and $\Gamma_\beta$ denote projection operators on the spin-isospin channels, and $c_{2\alpha\beta}$ are the adjustable free parameters. Again, this Lagrangian contains the effects of the pion Yukawa potential, but apart from that, all other short-range effects are modelled by the point-like contact interactions. This Lagrangian properly describes all NN scattering lengths, not only in the high-$L$ partial waves where the OPE potential is enough, but also in the low-$L$ partial waves.

Recently, ideas of the EFT for the NN scattering were followed further, by also adding to Lagrangian density (56) terms which contain six nucleon fields, and calculating the full energy dependence of phase shifts and mixing parameters in all partial waves [20]. The resulting effective Lagrangian density has many adjustable parameters, but the number of these parameters is comparable to that used in the parameterization of Lagrangian by heavy mesons. Also the description of the NN scattering data is of a comparable quality, i.e., very good. This shows that the ideas of the EFT really work; namely, it is not important which physical mechanism is used to model the short-range effects – a purely phenomenological mechanism is equally good. Our knowledge of these short-range effects can be summarized in a form of a certain number of constants that have the meaning of the multipole moments discussed above. Of course, it would be fantastically interesting to calculate these constants from the basic theory (QCD), but the description of low-energy nuclear phenomena requires only that these constants be known, while the whole complication of the vacuum, pion, and nucleon states does not enter the game.

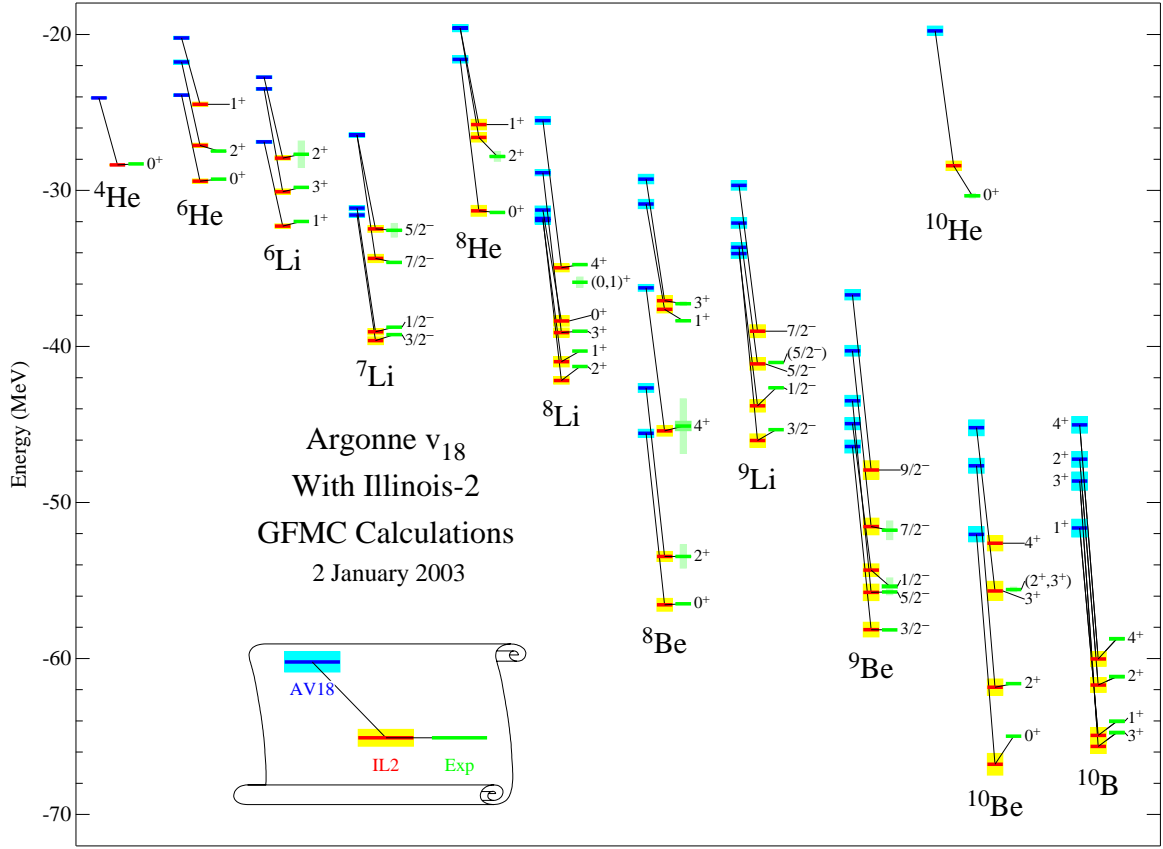

Figure 7: Results of the GFMC calculations for $A\leq 10$ nuclei. (Picture courtesy: S.C. Pieper, Argonne National Laboratory.)

## 3.4 Light Nuclei

Let us finish this Section with a brief discussion of the *ab initio* calculations for light nuclei. By using the Green Function Monte Carlo (GFMC) methods, one is able to determine binding energies, and energies of low-lying excited states, for systems containing up to $A$=10 nucleons [21, 22]. When the Argonne $v_{18}$ NN potential is used in such calculations, all light nuclei come out significantly underbound, see Fig. 7. The most plausible reason for such a discrepancy is the absence of the three-body NNN interaction, which is, as discussed in Sec. 3.1, expected to be a natural component of the force, and incorporates the polarization effects of the quark-gluon structure of the nucleons. Unfortunately, the scattering data only give us information on the binary NN component, and the three-body piece has to be postulated independently. When the Illinois NNN interaction [23] is added, the GFMC calculations reproduce properties of light nuclei with a very good precision (Fig. 7).

# 4 MANY-NUCLEON SYSTEMS

Let us now consider a system of many nucleons combined together within one composite object. We know that such composite particles (called nuclei or nuclides, as you know) exist in Nature. There exist exactly 253 species of stable nuclei.[1] About 2500 other ones have been synthesized in laboratories – they decay by different processes, like electron, positron, proton, or neutron emission, or by fission, i.e., by splitting into two lighter nuclei (including the case when one of the lighter nuclei is the $^4$He nucleus, called the $\alpha$ particle). According to theoretical predictions, there probably exist another 3000 nuclei, not synthesized yet, that are stable with respect to nucleon emission. At present, their synthesis, investigation, and description is at the centre of interest of nuclear structure physicists, and most of the lectures presented during this Summer School were devoted precisely to this subject.

Nuclei are fascinating objects. They are fermionic systems that exhibit single-particle (s.p.) and collective features at the same scale. Apart from very light ones, they contain too many constituents for an application of exact methods, and too few constituents for an application of statistical methods. Their elementary modes of excitation can, nevertheless, be very well defined based on using quasi-constituents and/or effective interactions.

## 4.1 General Discussion of the Nuclear Many-Body Problem

We begin our discussion of many-nucleon systems by (again) identifying the most important degrees of freedom and writing down the relevant Hamiltonian. Contrary to methods used at a finer level (quarks and gluons) we use here the Hamiltonian picture instead of the Lagrangian density; this is so because most of the analysis can be done in the framework of the standard quantum mechanics, without necessity of applying methods of the quantum field theory. Nevertheless, we shall express our many-body Hamiltonian in the language of the fermion creation and annihilation operators, which is very convenient in any theory that involves many identical particles obeying specific exchange symmetries.

In order to simplify the discussion we disregard the three-body NNN piece of the interaction between the nucleons, and thus the most general Hamiltonian of a many-nucleon system can be written as,

$$\hat{H} = T_{xy} a_x^+ a_y + \tfrac{1}{4} V_{xyx'y'} a_x^+ a_y^+ a_{y'} a_{x'}, \tag{57}$$

where $x \equiv (\boldsymbol{x}, \sigma, \tau)$, $x' \equiv (\boldsymbol{x}', \sigma', \tau')$, etc., are the space-spin-isospin variables, and the summation-integration $\int d^3\boldsymbol{r} \sum_{\sigma\tau}$ is implied for every pair of repeated indices. Following the standard notation, we put the space-spin-isospin arguments as indices of the kinetic energy, $T_{xy}$, potential energy, $V_{xyx'y'}$, and the creation $a_x^+$ and annihilation $a_y$ operators. We assume that the two-body potential energy operator is antisymmetrized, $V_{xyx'y'} = -V_{xyy'x'}$.

We can now estimate the order of complication involved in a many-nucleon system. Let us assume that fields $a_x^+$ (i.e., the s.p. wave functions) have to be known at about $M \simeq 10^4$ space-spin-isospin points. The estimate involves, say, about 20 points of a 1 fm lattice in each of the three spatial direction, and four spin-isospin components. The 1 fm lattice may seem to be grossly insufficient to describe a system where a typical s.p. kinetic energy $E_k$ is of the order of 50 MeV, and thus involves typical s.p. momenta of nucleons $k = \sqrt{2m_N E_k} \simeq 300$ MeV $\simeq 1.3\,\text{fm}^{-1}$ $\simeq (0.7\,\text{fm})^{-1}$. However, typical scale at which total densities of nucleons vary in a nucleus, are of the order of 2–3 fm, so the 1 fm lattice is a barely sufficient, but fair compromise to describe a system having the total size (including the asymptotic peripheral region) of at least 20 fm.

---
[1]Including several ones that live billions of years, and thus appear naturally on the Earth.

The fermion Fock space, i.e., the complete Hilbert space that is relevant to describe a system of many identical fermions, has the dimensionality of $D = \binom{M}{A}$, which is equal to the number of ways $A$ fermions can be distributed on $M$ sites. For the $A=10$ systems, which at present can still be treated within the GFMC method, Sect. 3.4, we thus obtain $D \simeq 10^{33}$. On the one hand, this number illustrates the power of the existing theoretical descriptions; on the other hand, it explains why it is so difficult to go any further. For example, for a heavy $A=200$ nucleus, the dimensionality reaches $10^{425}$. Therefore, it is neither conceivable nor sensible to envisage any exact methods for heavy nuclei.

One has to bear, however, in mind that the physics of a heavy nucleus does not really require such a detailed knowledge of any of its states. To see this, let us consider the energy of an arbitrary state $|\Psi\rangle$ as given by the average value of the Hamiltonian,

$$E = \langle \Psi | \hat{H} | \Psi \rangle = T_{xy} \rho_{yx} + \tfrac{1}{4} V_{xyx'y'} \rho_{x'y'xy}, \tag{58}$$

where the one- and two-body density matrices are defined as

$$\rho_{yx} = \langle \Psi | a_x^+ a_y | \Psi \rangle, \tag{59}$$
$$\rho_{x'y'xy} = \langle \Psi | a_x^+ a_y^+ a_{y'} a_{x'} | \Psi \rangle. \tag{60}$$

Both density matrices are Hermitian, $\rho_{yx} = \rho_{xy}^*$ and $\rho_{x'y'xy} = \rho_{xyx'y'}^*$, and the (fermion) two-body density matrix is antisymmetric with respect to exchanging its first two, or last two arguments, $\rho_{x'y'xy} = -\rho_{y'x'xy} = -\rho_{x'y'yx}$. Hence the total energy of an *arbitrary* many-fermion state, described by two-body interactions, is determined by $M^2 + (M(M-1)/2)^2 \simeq 10^{16}$ real parameters for $M \simeq 10^4$ (independently of $A$). Even when the three-body interactions are taken into account, this number grows "only" to $10^{24}$. This shows explicitly, that the information contained in a many-fermion nuclear state is, in fact, much smaller than the total dimensionality of the Hilbert space, or in other words, only very specific states from this Hilbert space are relevant.

Unfortunately, the presented counting rules, based on the analysis of density matrices, do not help in obtaining practical solutions for many-body problems. The reason for that is the never-solved $N$-representability problem [24, 25], namely, the question: which of the four-index matrices are two-body density matrices of many-fermion states, and which are not. Indiscriminate variation of Eq. (58) over the density matrices (to look for the ground state) is, therefore, inappropriate. Hence, we are back to square one, i.e., we have to anyhow consider the full Hilbert space to look for correct many-fermion states, even if we know that this constitutes an enormous waste of effort. New bright ideas to solve the $N$-representability problem in nuclear-physics context are very much needed. Before this is achieved, we are bound to look for methods judiciously reducing the dimensionality of the many-body problems. There are two main avenues to do so, which we briefly describe in the next two Sections.

## 4.2 Effective Interactions (I)

We saw that the crucial element of the dimensionality is the number of space-spin-isospin points needed to describe basic fields $a_x^+$. Therefore, we have to use methods that lead to fields as slowly varying in function of position, as it is possible. In this respect, region of the phase space that corresponds to pairs of nucleons getting near one another, is particularly cumbersome, because the wave functions must vary rapidly there, in order to become very small within the radius of the strong repulsion, cf. Fig. 5 above. In the past, very powerful technics have been developed to treat these hard-core effects. They are based on replacing the real NN interaction $V_{xyx'y'}$ by the effective interaction $G_{xyx'y'}$ that fulfills the following condition

$$\sum\!\!\!\!\!\!\int \mathrm{d}x'\mathrm{d}y'\, V_{xyx'y'}\Psi_{ij}(x',y') = \sum\!\!\!\!\!\!\int \mathrm{d}x'\mathrm{d}y'\, G_{xyx'y'} \left[\tfrac{\phi_i(x')\phi_j(y') - \phi_i(y')\phi_j(x')}{\sqrt{2}}\right], \tag{61}$$

where the sum-integrals are performed over $x'$ and $y'$.

The two-body wave function in the square brackets on the r.h.s. is the independent-particle, or product wave function, built as the antisymmetrized product of two s.p. wave functions, $\phi_i(x)$ and $\phi_j(x)$, characterized by quantum numbers $i$ and $j$. The two-body wave function on the l.h.s., $\Psi_{ij}(x',y')$, is a wave function correlated at the short range; it is very small within the region of the hard core. So the real NN interaction, when acting on the correlated wave function, gives a finite result, because the wave function is very small in the region where the repulsion is vary large. On the other hand, the antisymmetrized product wave function is never small around $x'{=}y'$ (although it vanishes at $x'{=}y'$), and hence the effective interaction fulfilling (61) *has no hard core*. Condition (61) defines, therefore, the effective interaction that can be used in the space of uncorrelated Slater determinants. The whole procedure can be put on firm grounds in the framework of the perturbation expansion, when partial sums of infinite classes of diagrams are performed, but this is beyond the scope of the present lectures. We only mention that within such a formalism, the effective interaction is obtained by solving the Bethe-Goldstone equation [26].

The effective interaction should, in principle, depend on the s.p. states $\phi_i(x)$ and $\phi_j(x)$ for which the Bethe-Goldstone equation is solved. For example, the effective interaction in an infinite nuclear matter, where the s.p. wave functions are plane waves, can be different than that in a finite nucleus. In the past, there were many calculations pertaining to the first case, while the second (and more interesting) situation was successfully addressed only very recently [27, 28].

On a phenomenological level, one can postulate simple forms of interactions and use them as models of such difficult-to-derive effective interactions. Such a route was adopted by Gogny [29], who postulated the simple local interaction

$$\tilde{G}_{xyx'y'} = \delta(\boldsymbol{x} - \boldsymbol{x}')\delta(\boldsymbol{y} - \boldsymbol{y}')G(x,y), \tag{62}$$

where the tilde denotes a non-antisymmetrized matrix element ($G_{xyx'y'} = \tilde{G}_{xyx'y'} - \tilde{G}_{xyy'x'}$), in the form of a sum of two Gaussians, plus a zero-range, density dependent part,[2]

$$\begin{aligned} G(x,y) &= \sum_{i=1,2} e^{-(\boldsymbol{x}-\boldsymbol{y})^2/\mu_i^2} \times (W_i + B_i P_\sigma - H_i P_\tau - M_i P_\sigma P_\tau) \\ &+ t_3(1 + P_\sigma)\delta(\boldsymbol{x} - \boldsymbol{y})\rho^{1/3}\left[\tfrac{1}{2}(\boldsymbol{x} + \boldsymbol{y})\right]. \end{aligned} \tag{63}$$

In this Equation, $P_\sigma{=}\tfrac{1}{2}(1{+}\boldsymbol{\sigma}_1\cdot\boldsymbol{\sigma}_2)$ and $P_\tau{=}\tfrac{1}{2}(1{+}\vec{\tau}_1\circ\vec{\tau}_2)$ are, respectively, the spin and isospin exchange operators of particles 1 and 2, $\rho(\boldsymbol{r})$ is the total density of the system at point $\boldsymbol{r}$, and $\mu_i$, $W_i$, $B_i$, $H_i$, $M_i$, and $t_3$ are parameters.

In Fig. 8, we compare the real $n$-$n$ interaction (Argonne $v_{18}$) with the effective Gogny interaction (the D1 parametrization [29, 30]) in the $L{=}0$ channels, i.e., in the $^3S_1$ channel ($P_\sigma{=}1$ and $P_\tau{=}{-}1$) and $^1S_0$ channel ($P_\sigma{=}{-}1$ and $P_\tau{=}1$). It is clear that real and effective interactions are very different near $r$=0. The zero-range piece of the interaction acts only in the $^3S_1$ channel; in Fig. 8 it is represented by the green arrow at $r$=0. One should keep in mind that the Gogny interaction is meant to represent the effective interaction, and hence it can only act on the product wave functions. In particular, an attempt to solve exactly, e.g., the two-body (deuteron) problem goes beyond the range of applicability of the effective interaction. The Gogny interaction is mostly used within the mean-field approximation that we discuss in more detail in the Sec. 4.4 below.

---

[2]We omit the spin-orbit term for simplicity.

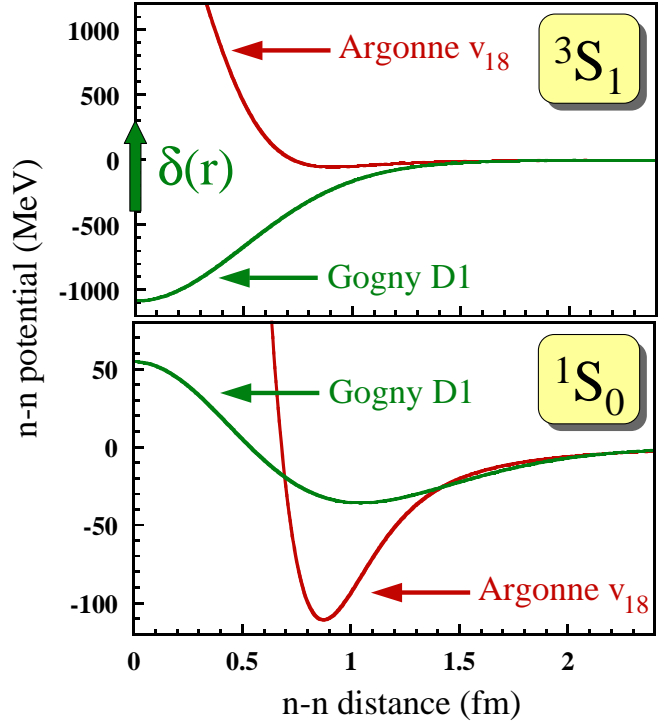

Figure 8: Comparison of the Gogny and Argonne $v_{18}$ $n$-$n$ potentials in the $^3S_1$ (top) and $^1S_0$ (bottom) channels. Note very different scales of the top and bottom panels.

## 4.3 Effective Interactions (II)

To a certain extent, a way out from the explosion of dimensionality, discussed in Sec. 4.1, may consist in using a better single-particle space. Instead of parametrizing fields $a_x^+$ by space-spin-isospin points $x$, one can use a parametrization by the shell-model orbitals $\phi_i(x)$ that are active near the Fermi surface of a given nucleus, i.e., by fields

$$a_i^+ = \sum \!\!\!\!\!\!\int \mathrm{d}x \, \phi_i(x) a_x^+. \qquad (64)$$

When a complete set of orbitals is used, the descriptions in terms of creation operators $a_i^+$ and $a_x^+$ are equivalent. However, one can also attempt a drastic reduction of the set $a_i^+$ to a finite number, $i$=1...$M$, of "most important" orbitals, similarly as we have been previously using finite sets of the space-spin-isospin points instead of continuous variables.

The reduction is now not a mere question of discretizing continuous fields, but involves a serious limitation of the Hilbert space. In quantum mechanics one can always split the Hilbert space into two subspaces, $|\Psi\rangle = P|\Psi\rangle + Q|\Psi\rangle$, where $P$ and $Q$ are projection operators such that $P+Q=1$. Then, the Schrödinger equation $H|\Psi\rangle = E|\Psi\rangle$ is strictly equivalent to the following 2×2 matrix of equations,

$$\begin{pmatrix} PHP & PHQ \\ QHP & QHQ \end{pmatrix} \begin{pmatrix} P|\Psi\rangle \\ Q|\Psi\rangle \end{pmatrix} = E \begin{pmatrix} P|\Psi\rangle \\ Q|\Psi\rangle \end{pmatrix} \quad . \tag{65}$$

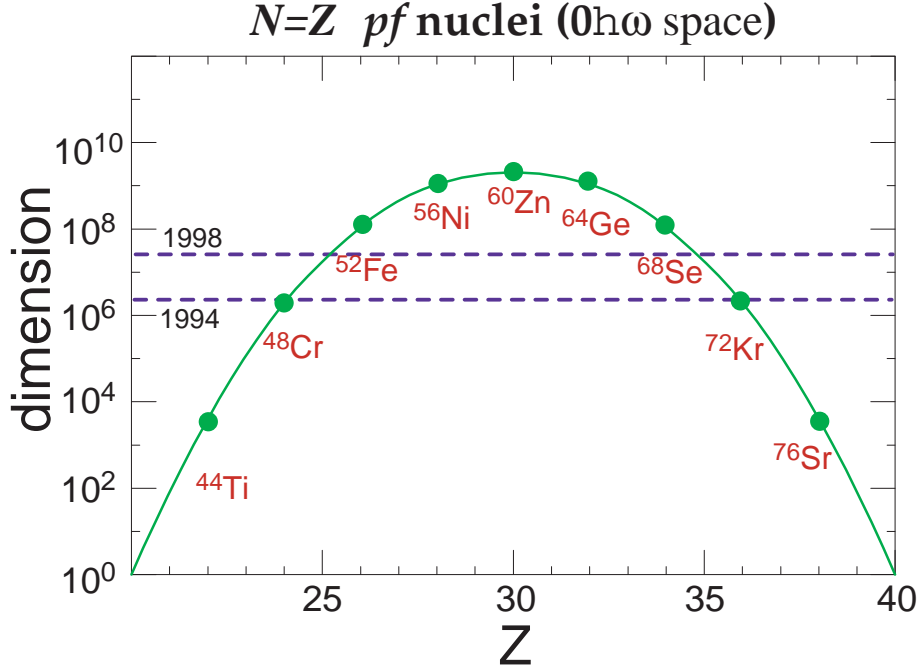

Figure 9: Dimension of the shell-model space for calculations of $N=Z$ nuclei within the $pf$ space. (Picture courtesy: W. Nazarewicz, ORNL/University of Tennessee/Warsaw University.) From `http://www-highspin.phys.utk.edu/~witek/`.

Using the second equation, one can now formally express the "excluded" component, $|\Psi_Q\rangle \equiv Q|\Psi\rangle$, of the wave function by the "kept" component, $|\Psi_P\rangle \equiv P|\Psi\rangle$, i.e.,

$$|\Psi_Q\rangle = \frac{1}{E - QH} QH|\Psi_P\rangle, \tag{66}$$

and put it back into the first equation. This gives the Schrödinger equation reduced to the "kept" Hilbert space,

$$H_{\text{eff}}|\Psi_P\rangle = E|\Psi_P\rangle, \tag{67}$$

where the effective Hamiltonian $H_{\text{eff}}$ is given by the Bloch-Horowitz equation [31],

$$H_{\text{eff}} = H + H \frac{1}{E - QH} QH \quad . \tag{68}$$

The main questions is, of course, whether the Bloch-Horowitz effective interaction, $V_{\text{eff}} = H_{\text{eff}} - T$, can be replaced by a simple phenomenological interaction, and used to describe real systems. In particular, when a two-body, energy-independent interaction is postulated in a very small phase space, one obtains the shell model, which is successfully used since many years in nuclear structure physics.

In order to illustrate the dimensions of the shell-model Hilbert space, in Fig. 9 we show the numbers of many-fermion states that are obtained when states in $N=Z$ medium heavy nuclei are described within the $pf$ space (20 s.p. states for protons and 20 for neutrons). Currently, complete solutions for the $pf$ space become available, i.e., dimensions of the order of $10^9$ can effectively be treated. Progress in this domain closely follows the progress in size and speed of computers, i.e., one order of magnitude is gained in about every two-three years. We shell not discuss these methods in any more detail, because dedicated lectures have been presented on this subject during the Summer School.

## 4.4 Hartree-Fock method

The Hartree-Fock (HF) approach relies on assuming that the ground state of a many-fermion system can be uniquely characterized by the one-body density matrix (59). There are many ways of deriving the HF equations; the simplest one is to use the variational principle together with the following approximation of the two-body density matrix (60):

$$\rho_{x'y'xy} = \rho_{x'x}\rho_{y'y} - \rho_{x'y}\rho_{y'x}. \tag{69}$$

This equation expresses the two-body density matrix by the one-body density matrix, and hence the total energy (58) becomes a functional of the one-body density matrix only,

$$\begin{aligned} E_{\text{HF}} &= T_{xy}\rho_{yx} + \tfrac{1}{4}G_{xyx'y'}\left(\rho_{x'x}\rho_{y'y} - \rho_{x'y}\rho_{y'x}\right) \\ &= T_{xy}\rho_{yx} + \tfrac{1}{2}\Gamma_{xx'}\rho_{x'x} = \tfrac{1}{2}\left(T_{xy} + h_{xy}\right)\rho_{yx}, \end{aligned} \tag{70}$$

for

$$\begin{aligned} \Gamma_{xx'} &= G_{xyx'y'}\rho_{y'y} &&\Longleftarrow\quad \text{HF potential,} \tag{71}\\ h_{xy} &= T_{xy} + \Gamma_{xy} &&\Longleftarrow\quad \text{HF Hamiltonian.} \tag{72} \end{aligned}$$

By minimizing the HF energy (70) with respect to the one-body density matrix, one obtains

$$h_{xy}\rho_{yz} - \rho_{xy}h_{yz} = 0 \quad\Longleftarrow\quad \text{HF equation,} \tag{73}$$

which is usually solved by finding the HF s.p. orbitals that diagonalize the HF Hamiltonian (72),

$$\sum\!\!\!\!\!\!\int \mathrm{d}y\, h_{xy}\phi_i(y) = \epsilon_i\phi_i(x), \tag{74}$$

and then constructing the one-body density matrix from these orbitals:

$$\rho_{xy} = \sum_{i\in\text{occ}} \phi_i(x)\phi_i^*(y). \tag{75}$$

Equations (74) and (75) guarantee that the HF condition (73) is fulfilled (because $h_{xy}$ and $\rho_{xy}$ are then diagonal in the common basis), so the HF solution is found whenever, for a given set of occupied orbitals, $i \in$ occ, the density matrix self-consistently reproduces the HF potential (71).

From Eq. (75) it is clear that not the real interaction $V_{xyx'y'}$, but the effective interaction $G_{xyx'y'}$, must be used in the HF method. Indeed, when the density-matrix (75) is inserted in the expression for the HF energy (70), one recovers the action of the effective interaction on the two-body product wave functions (61). It is now obvious that the determination of the effective interaction must be coupled to the solution of the HF equations, and performed self-consistently. Namely, for a given effective interaction one solves the HF equations, and the obtained HF orbitals (74) are in turn used in the Bethe-Goldstone equation to find effective interaction. Such a doubly self-consistent procedure is called the Brueckner-Hartree-Fock method.

Modern understanding of the HF approximation is not directly based on the variational method applied to Slater determinants. Certainly, the basic approximation for the two-body density matrix (69) is an exact result for a Slater determinant, but the key element of the approach is expression (70), which states that the ground-state energy can be approximated by a functional of the one-body density matrix.

## 4.5 Conserved and Broken Symmetries

Representation of many-fermion states by density matrices (59) and (60), and the HF approximation of the two-body density matrix (69), allow us to give a precise definition of what one really means by conserved and broken symmetries in many-body systems. Moreover, it also links the spontaneous symmetry breaking mechanism to a description of correlations.

Consider a unitary symmetry operator $\hat{P}$ such that

$$\hat{P}^+ a_x \hat{P} = P_{xy} a_y \quad , \quad \hat{P}^+ a_x^+ \hat{P} = P_{xy}^* a_y^+ \tag{76}$$

and

$$P_{xx'}^+ T_{x'y'} P_{y'y} = T_{xy} \quad , \quad P_{zz'}^+ P_{tt'}^+ G_{z't'x'y'} P_{x'x} P_{y'y} = G_{ztxy}. \tag{77}$$

Equations (76) and (77) are equivalent to the symmetry condition $[\hat{H}, \hat{P}]=0$ obeyed by Hamiltonian (57). Symmetry operator $\hat{P}$ acts in the fermion Fock space by mixing elementary fields $a_y^+$ with the integral kernel $P_{xy}$ (remember that the sum-integral $\sum\!\!\!\!\!\int dy$ is implied for every repeated index). All the most interesting symmetries act in this way – they can be represented as exponents of one-body symmetry generators, i.e., $\hat{P}$ can be any one of the following:

1° translational symmetry,
$$\hat{P} = \exp\left(i\boldsymbol{r}_0 \cdot \hat{\boldsymbol{P}}\right), \tag{78}$$
where $\hat{\boldsymbol{P}} = \sum_{i=1}^A \boldsymbol{p}_i$ is the total linear momentum operator, and $\boldsymbol{r}_0$ is the shift vector.

2° rotational symmetry,
$$\hat{P} = \exp\left(i\boldsymbol{\alpha}_0 \cdot \hat{\boldsymbol{I}}\right), \tag{79}$$
where $\hat{\boldsymbol{I}} = \sum_{i=1}^A \boldsymbol{j}_i$ is the total angular momentum operator, and $\boldsymbol{\alpha}_0$ is the rotation angle.

3° isospin symmetry,
$$\hat{P} = \exp\left(i\vec{\alpha}_0 \circ \hat{\vec{T}}\right), \tag{80}$$
where $\hat{\vec{T}} = \frac{1}{2}\sum_{i=1}^A \vec{\tau}_i$ is the total isospin operator, and $\vec{\alpha}_0$ is the iso-rotation angle.

4° particle-number symmetry,
$$\hat{P} = \exp\left(i\phi_0 \hat{N}\right), \tag{81}$$
where $\hat{N} = \sum\!\!\!\!\!\int dx\, a_x^+ a_x$ is the total particle number operator, and $\phi_0$ is the gauge angle.

5° inversion (parity) symmetry,
$$\hat{P} = \prod_{i=1}^A \hat{\pi}_i, \tag{82}$$
where $\hat{\pi}_i$ is the inversion operator for the $i$th particle.

6° time-reversal symmetry.
$$\hat{P} = \exp\left(-i\pi \hat{S}_y\right) \hat{K}, \tag{83}$$

where $\hat{S}_y = \frac{1}{2}\sum_{i=1}^{A} \sigma_{iy}$ is the $y$ component of the total spin operator, and $\hat{K}$ is the complex conjugation operator in spatial representation.

There can also be terms in the Hamiltonian that *explicitly* break some of the above symmetries (e.g., the Coulomb interaction explicitly breaks the isospin symmetry), but we disregard them for simplicity.

Let us begin with the simplest case, namely, let $\hat{P}$ be the parity symmetry (82). In this case, the integral kernel reads $P_{xy} \equiv \delta(\boldsymbol{x}+\boldsymbol{y})$, and is, of course, independent of spin and isospin. For a parity-invariant interaction, Eq. (77), the exact energy of an arbitrary state $|\Psi\rangle$, Eq. (58), depends only on the scalar parts (in this case, the parity invariant parts) of the one- and two-body density matrices, i.e.,
$$E = T_{xy}\rho_{yx}^{(+)} + \tfrac{1}{4}G_{xyzt}\rho_{ztxy}^{(+)}, \tag{84}$$
for
$$\rho_{yx}^{(\pm)} = \tfrac{1}{2}\left(P_{yy'}\rho_{y'x'}P_{x'x}^+ \pm \rho_{yx}\right), \tag{85}$$
$$\rho_{ztxy}^{(\pm)} = \tfrac{1}{2}\left(P_{zz'}P_{tt'}\rho_{z't'x'y'}P_{x'x}^+ P_{y'y}^+ \pm \rho_{ztxy}\right). \tag{86}$$

Within the HF approximation (69), we may have two classes of solutions:

- symmetry-conserving solution:
$$\rho_{xy} = \rho_{xy}^{(+)}, \tag{87}$$
$$\rho_{ztxy}^{(+)} = \rho_{zx}^{(+)}\rho_{ty}^{(+)} - \rho_{zy}^{(+)}\rho_{tx}^{(+)}, \tag{88}$$

- symmetry-breaking solution:
$$\rho_{xy} = \rho_{xy}^{(+)} + \rho_{xy}^{(-)}, \tag{89}$$
$$\rho_{ztxy}^{(+)} = \rho_{zx}^{(+)}\rho_{ty}^{(+)} - \rho_{zy}^{(+)}\rho_{tx}^{(+)} + \rho_{zx}^{(-)}\rho_{ty}^{(-)} - \rho_{zy}^{(-)}\rho_{tx}^{(-)}. \tag{90}$$

In the case of the broken symmetry, neither of the density matrices is invariant with respect to the symmetry operator. However, the symmetry breaking part of the one-body density matrix $\rho_{xy}^{(-)}$ enters the HF energy (84) only through the two-body interaction energy. Moreover, the symmetry-projected two-body density matrix (90) *does not obey* the HF condition (69). In other words, the symmetry-breaking part of the one-body density matrix gives a *correlation* term of the two-body density matrix. Symmetry breaking is, therefore, a reflection of correlations beyond HF, taken into account with respect to the symmetry-conserving HF method.

One can also say that the symmetry-breaking part $\rho_{xy}^{(-)}$ constitutes an additional set of variational parameters, which become allowed when a larger class of the one-body density matrices (beyond symmetry conservation) is considered. As in every variational procedure, a larger variational class may lead (sometimes) to lower energies. Whether it does, depends on the specific case, and in particular on the type of the two-body interaction. It is obvious, that one can gain energy by breaking symmetry only if the appropriate correlation energy is negative, i.e., when the last two terms of the two-body density matrix, $\rho_{zx}^{(-)}\rho_{ty}^{(-)} - \rho_{zy}^{(-)}\rho_{tx}^{(-)}$, give a negative contribution when averaged with the two-body effective interaction $G_{xyzt}$.

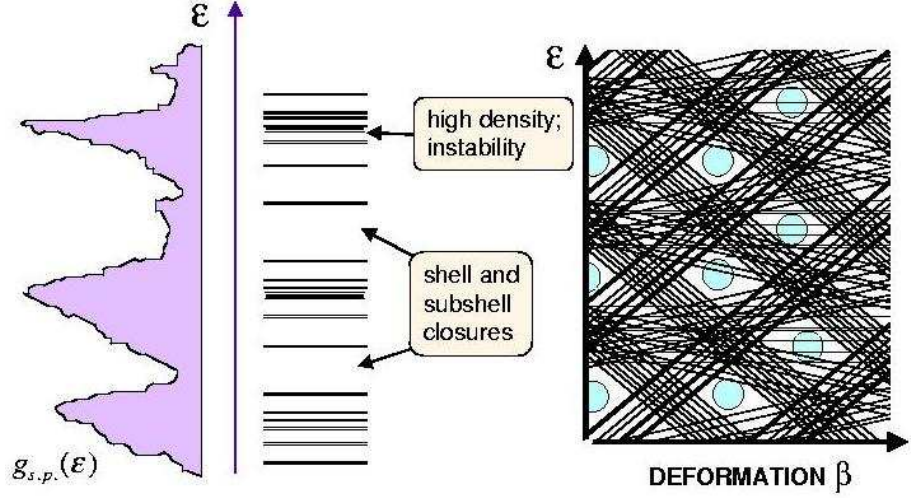

Figure 10: Schematic illustration of the s.p. level density (left), corresponding to the s.p. spectrum of a deformed nucleus (centre). The right panel shows the evolution of the spectrum with nuclear deformation. (Picture courtesy: W. Nazarewicz, ORNL/University of Tennessee/Warsaw University.) From `http://www-highspin.phys.utk.edu/~witek/`.

Within such an approach to the symmetry breaking, one does not, in fact, break any symmetry of the exact wave function. Indeed, the density matrices, $\rho_{xy}^{(+)}$ and $\rho_{ztxy}^{(+)}$ that are "active" in the total energy do conserve the symmetry. We should also use these density matrices to calculate all other observables for the symmetry-broken (correlated) solution of the HF equations.

Let us now give results of an analogous analysis for the case of deformed nuclei, i.e., for the case of broken rotational symmetry (79). For axial shapes we then have the following density matrices,

$$\rho_{xy} = \sum_J \rho_{xy}^{(J)}, \tag{91}$$

$$\rho_{ztxy}^{(0)} = \sum_J \left(\rho_{zx}^{(J)} \times \rho_{ty}^{(J)}\right)_0 - \sum_J \left(\rho_{zy}^{(J)} \times \rho_{tx}^{(J)}\right)_0, \tag{92}$$

and the total HF energy,

$$E = T_{xy}\rho_{yx}^{(0)} + \tfrac{1}{4}G_{xyzt}\rho_{ztxy}^{(0)}, \tag{93}$$

that depends only on the scalar (J=0) parts of the density matrices. On the other hand, the broken-symmetry one-body density matrix is the sum of components $\rho_{xy}^{(J)}$ that transform as irreducible rotational tensors of rank $J$. In the scalar two-body density matrix (92), these components are coupled to J=0, and every such a term defines the multipole correlation energy of rank $J$. It is now obvious that the broken-symmetry solution becomes the ground state for interactions that have appropriately strong multipole-multipole terms (see Refs. [32, 33] for numerical results in heavy nuclei).

Without going into detailed discussion of the multipole-multipole decomposition of effective interactions, we may easily tell in which nuclei the rotational symmetry is broken and deformation appears. A schematic diagram presented in the right panel of Fig. 10 shows the evolution of the s.p. spectrum with nuclear deformation, i.e., the dependence of eigenvalues of the mean-field Hamiltonian having the shape characterized by the deformation parameter $\beta$. In such a spectrum, some s.p. levels go down, and other go up in energy, and at specific deformations there appear in the spectrum larger or smaller gaps. When the particles are filling the lowest levels up to certain energy (prescribed by the number of particles), the last occupied level may appear either below or above the gap. This leads respectively to a decrease or an increase of the total energy. The overall density of s.p. levels at the Fermi surface determines, therefore, the total energy of the system. In other words, a system having a given number of particles adopts the shape at which the last occupied level is below a large gap. Therefore, nuclei that correspond to magic particle numbers are spherical (large gaps appear at spherical shape) and the rotational symmetry is conserved, while nuclei with particle numbers between the magic gaps (the so-called open-shell nuclei) choose non-zero deformed ground states corresponding to broken rotational symmetry.

## 4.6 Local Density Approximation

Approximation of the many-body energy (58) by a functional of the one-body density matrix (70) can be further simplified in the coordinate representation. Namely, it appears that the HF density matrix (75) influences the energy mostly through the local density [34, 35, 36]. This observation defines the local density approximation (LDA).

Neglecting for simplicity the spin-isospin degrees of freedom, we can write the interaction energy [the second term in Eq. (70)] in the form

$$E_{\text{int}} = \tfrac{1}{2} \int \mathrm{d}^3\boldsymbol{x}\,\mathrm{d}^3\boldsymbol{y}\,\mathrm{d}^3\boldsymbol{x}'\mathrm{d}^3\boldsymbol{y}'\,\tilde{G}_{\boldsymbol{xyx'y'}} \left(\rho_{\boldsymbol{x'x}}\rho_{\boldsymbol{y'y}} - \rho_{\boldsymbol{x'y}}\rho_{\boldsymbol{y'x}}\right). \tag{94}$$

For local effective interaction, the non-antisymmetrized matrix element $\tilde{G}_{\boldsymbol{xyx'y'}}$ is given by the potential $G(\boldsymbol{x},\boldsymbol{y})$,

$$\tilde{G}_{\boldsymbol{xyx'y'}} = \delta(\boldsymbol{x}-\boldsymbol{x}')\delta(\boldsymbol{y}-\boldsymbol{y}')G(\boldsymbol{x},\boldsymbol{y}), \tag{95}$$

and the interaction energy reads

$$E^{\text{int}} = \tfrac{1}{2} \int \mathrm{d}^3\boldsymbol{x}\,\mathrm{d}^3\boldsymbol{y}\,G(\boldsymbol{x},\boldsymbol{y})\left(\rho_{\boldsymbol{xx}}\rho_{\boldsymbol{yy}} - \rho_{\boldsymbol{xy}}\rho_{\boldsymbol{yx}}\right). \tag{96}$$

The first term (direct) depends only on the local density matrix (equal arguments), while the second term (exchange) involves the full one-body density matrix. Therefore, the local density plays a special role due to locality of the effective interaction.

It is therefore convenient to represent the one-body density matrix (59) in total and relative coordinates, i.e.,

$$\rho_{\boldsymbol{xy}} = \rho(\boldsymbol{R},\boldsymbol{r}), \tag{97}$$

where

$$\boldsymbol{R} = \tfrac{1}{2}(\boldsymbol{x}+\boldsymbol{y}) \quad\text{and}\quad \boldsymbol{r} = \boldsymbol{x}-\boldsymbol{y}. \tag{98}$$

Denoting the local density by single argument, $\rho(\boldsymbol{R}) = \rho_{\boldsymbol{xx}} = \rho(\boldsymbol{R},\boldsymbol{0})$, and noting that by translational invariance the potential $G(\boldsymbol{x},\boldsymbol{y}) = G(\boldsymbol{x}-\boldsymbol{y})$ depends only on the relative coordinate, we have

$$E^{\text{int}} = \tfrac{1}{2}\int \mathrm{d}^3\boldsymbol{R}\,\mathrm{d}^3\boldsymbol{r}\,G(\boldsymbol{r})\left[\rho(\boldsymbol{R}+\tfrac{1}{2}\boldsymbol{r})\rho(\boldsymbol{R}-\tfrac{1}{2}\boldsymbol{r}) - \rho(\boldsymbol{R},\boldsymbol{r})\rho(\boldsymbol{R},-\boldsymbol{r})\right]. \tag{99}$$

We see that direct and exchange terms, $E^{\text{int}} = E^{\text{int}}_{\text{dir}} + E^{\text{int}}_{\text{exch}}$, have markedly different dependence on the density matrix, and thus have to be treated separately.

In the direct term, we can use the fact that the range of the effective force is smaller than the typical distance at which the density changes. Indeed, the nuclear density is almost constant inside the nucleus, and then falls down to zero within the region called the nuclear surface, which has a typical width of about 3 fm. Hence, within the range of interaction, and for the purpose of evaluating the direct interaction energy, the density can be approximated by the quadratic expansion,

$$\rho(\boldsymbol{R} \pm \tfrac{1}{2}\boldsymbol{r}) = \rho(\boldsymbol{R}) \pm \tfrac{1}{2}r^i\nabla_i\rho(\boldsymbol{R}) + \tfrac{1}{8}r^ir^j\nabla_i\nabla_j\rho(\boldsymbol{R}) + \ldots \tag{100}$$

and

$$\rho(\boldsymbol{R}+\tfrac{1}{2}\boldsymbol{r})\rho(\boldsymbol{R}-\tfrac{1}{2}\boldsymbol{r}) = \rho^2(\boldsymbol{R}) + \tfrac{1}{4}r^ir^j\Big(\rho(\boldsymbol{R})\nabla_i\nabla_j\rho(\boldsymbol{R}) - [\nabla_i\rho(\boldsymbol{R})][\nabla_j\rho(\boldsymbol{R})]\Big) + \ldots, \tag{101}$$

where $\nabla_i = \partial/\partial R^i$. When inserted into Eq. (99), this expansion gives [for scalar interactions $G(\boldsymbol{r}) = G(|\boldsymbol{r}|) = G(r)$] the **direct** interaction energy:

$$E_{\rm dir}^{\rm int} = \tfrac{1}{2}\int d^3\boldsymbol{R}\Big[G_0\rho^2 + \tfrac{1}{4}G_2\big(\rho\Delta\rho - (\boldsymbol{\nabla}\rho)^2\big)\Big] + \ldots, \tag{102}$$

where coupling constants $G_0$ and $G_2$ are given by the moments of the interaction:

$$G_0 = 4\pi\int dr\, r^2 G(r) \quad \text{and} \quad G_2 = \tfrac{4}{3}\pi\int dr\, r^4 G(r). \tag{103}$$

In the exchange term, the situation is entirely different, because here the range of interaction matters in the non-local, relative direction $\boldsymbol{r}$. In order to get a feeling what are the properties of the one-body density matrix in this direction, we can calculate it for infinite matter,

$$\rho_{\boldsymbol{xy}} = \int_{|\boldsymbol{k}|<k_F} d^3\boldsymbol{k}\, \frac{\exp(i\boldsymbol{k}\cdot\boldsymbol{x})}{\sqrt{8\pi^3}}\frac{\exp(-i\boldsymbol{k}\cdot\boldsymbol{y})}{\sqrt{8\pi^3}}, \tag{104}$$

where the s.p. wave functions (plane waves) are integrated within the Fermi sphere of momenta $|\boldsymbol{k}| < k_F$. Obviously, $\rho_{\boldsymbol{xy}}$ depends only on the relative coordinate, i.e.,

$$\rho(\boldsymbol{R},\boldsymbol{r}) = \frac{1}{2\pi^2 r}\int_0^{k_F} dk\, k\sin(kr) = \frac{k_F^3}{6\pi^2}\left[3\frac{\sin(k_F r) - k_F r\cos(k_F r)}{(k_F r)^3}\right] = \frac{k_F^3}{6\pi^2}\left[3\frac{j_1(k_F r)}{k_F r}\right]. \tag{105}$$

Function in square parentheses equals 1 at $r=0$, and has the first zero at $r \simeq 4.4934/k_F \simeq 3\,{\rm fm}$, i.e., in the non-local direction the density varies on the same scale as it does in the local direction.

Therefore, the quadratic expansion of the density matrix in the relative variable

$$\rho(\boldsymbol{R},\pm\boldsymbol{r}) = \rho(\boldsymbol{R}) \pm r^i\partial_i\rho(\boldsymbol{R},\boldsymbol{r}) + \tfrac{1}{2}r^ir^j\partial_i\partial_j\rho(\boldsymbol{R},\boldsymbol{r}) + \ldots, \tag{106}$$

where derivatives $\partial_i = \partial/\partial r^i$ are always calculated at $r^i=0$, is, in principle, sufficient for the evaluation of the exchange interaction energy. However, we can improve it by introducing three universal functions of $r = |\boldsymbol{r}|$, $\pi_0(r)$, $\pi_1(r)$, and $\pi_2(r)$, which vanish at large $r$, i.e., we define the LDA by:

$$\rho(\boldsymbol{R},\pm\boldsymbol{r}) = \pi_0(r)\rho(\boldsymbol{R}) \pm \pi_1(r)r^i\partial_i\rho(\boldsymbol{R},\boldsymbol{r}) + \tfrac{1}{2}\pi_2(r)r^ir^j\partial_i\partial_j\rho(\boldsymbol{R},\boldsymbol{r}) + \ldots \tag{107}$$

Since for small $r$, Eq. (107) must be compatible with the Taylor expansion (106), the auxiliary functions must fulfill conditions at $r=0$,

$$\pi_0(0) = \pi_1(0) = \pi_2(0) = 1, \quad \pi_0'(0) = \pi_1'(0) = 0, \quad \text{and} \quad \pi_0''(0) = 0. \tag{108}$$

In order to conserve the local-gauge-invariance properties of the interaction energy [37], we also require that

$$\pi_1^2(r) = \pi_0(r)\pi_2(r). \tag{109}$$

The auxiliary functions $\pi_0(r)$ and $\pi_2(r)$ can be calculated *a posteriori*, to give the best possible approximation of a given density matrix $\rho(\boldsymbol{R}, \boldsymbol{r})$. However, they can also be estimated *a priori* by making momentum expansion around the Fermi momentum $k_F$. This gives the density-matrix expansion (DME) of Ref. [35], in which

$$\pi_0(r) = \frac{6j_1(k_F r) + 21j_3(k_F r)}{2k_F r} \quad \text{and} \quad \pi_2(r) = \frac{105 j_3(k_F r)}{(k_F r)^3}, \tag{110}$$

where $j_n(k_F r)$ are the spherical Bessel functions.

The term depending on the non-local density in the exchange integral (99) now reads

$$\rho(\boldsymbol{R}, \boldsymbol{r})\rho(\boldsymbol{R}, -\boldsymbol{r}) = \pi_0^2(r)\rho^2(\boldsymbol{R}) + \pi_0(r)\pi_2(r)r^i r^j \Big(\rho(\boldsymbol{R})\partial_i \partial_j \rho(\boldsymbol{R}, \boldsymbol{r}) - [\partial_i \rho(\boldsymbol{R}, \boldsymbol{r})][\partial_j \rho(\boldsymbol{R}, \boldsymbol{r})]\Big) + \ldots \tag{111}$$

and gives the **exchange** interaction energy:

$$E_{\text{exch}}^{\text{int}} = -\tfrac{1}{2} \int d^3 \boldsymbol{R} \Big[ G_0' \rho^2 + \tfrac{1}{4} G_2' \big(\rho \Delta \rho - 4(\rho \tau - \boldsymbol{j}^2)\big) \Big] + \ldots, \tag{112}$$

where coupling constants $G_0'$ and $G_2'$ are given by the following integrals of the interaction:

$$G_0' = 4\pi \int dr\, r^2 \pi_0^2(r) G(r) \quad \text{and} \quad G_2' = \tfrac{4}{3}\pi \int dr\, r^4 \pi_0(r)\pi_2(r) G(r). \tag{113}$$

The exchange interaction energy also depends on densities $\boldsymbol{j}$ (119) and $\tau$ (120) that we define below. It is obvious that when the pure Taylor expansion is used to approximate the density in the non-local direction, Eq. (106), i.e., for $\pi_0(r) = \pi_2(r) = 1$, the direct and exchange coupling constants are equal, $G_0' = G_0$ and $G_2' = G_2$.

Altogether, quadratic approximations to the one-body density matrix allow expressing the direct and exchange interaction energies as integrals of local energy density. Such energy density depends on the local density, on derivatives of the local density, and on several other densities that represent properties of the one-body density matrix in the non-local direction.

We should stress that the validity of the LDA depends on different scales involved in properties of nuclei. Namely, the scale of distances characterizing the ground-state one-body density matrix is significantly larger than the range of effective forces. Therefore, the LDA may apply only to selected, low-energy phenomena where the spatial structure of the density matrix is not very much affected.

Moreover, we see that the low-energy nuclear properties may depend on an extremely restricted set of properties of effective interactions. Within the LDA, only a few numbers [the coupling constants of Eqs. (103) and (113)] determine the energy density. This is entirely in the spirit of the effective field theory; separation of scales results in a transmission of a very limited information from one scale to another. Once this information (in our case – the coupling constants) is either evaluated, or fit to data, properties of the system can be properly calculated at the larger scale.

We also see that the coupling constants can be evaluated by assuming any effective interaction that has a *smaller* range than the physical range. In doing so, we can even go down to zero range, and nothing will change, provided we fix the parameters of the zero-range force so as to properly describe the moments of the force, Eqs. (103) and (113), and thus properly reproduce the coupling constants.

We can now proceed to the real world by putting back into our description the spin and isospin degrees of freedom. Based on the results above, we can first construct the most general set of local densities, with derivatives up to the second order taken into account, and then build the local energy density. The complete such construction has been performed only very recently [38]; it involves the full proton-neutron mixing and treats both the particle-hole and particle-particle channels of interaction.

We begin by writing the one-body density matrix (59) with all variables shown explicitly,

$$\rho_{\bm{x}\sigma\tau,\bm{y}\sigma'\tau'} = \langle\Psi|a^+_{\bm{y}\sigma'\tau'}a_{\bm{x}\sigma\tau}|\Psi\rangle, \tag{114}$$

and we define the densities in total and relative coordinates (97) as

$$\rho(\bm{R},\bm{r},\sigma\tau,\sigma'\tau') = \rho_{\bm{x}\sigma\tau,\bm{y}\sigma'\tau'}. \tag{115}$$

The spin-isospin components can now be separated,

$$\begin{aligned}\rho(\bm{R},\bm{r},\sigma\tau,\sigma'\tau') &= \tfrac{1}{4}\rho_0(\bm{R},\bm{r})\delta_{\sigma\sigma'}\delta_{\tau\tau'} + \tfrac{1}{4}\delta_{\sigma\sigma'}\vec{\rho}(\bm{R},\bm{r})\circ\vec{\tau}_{\tau\tau'} \\ &+ \tfrac{1}{4}\bm{s}_0(\bm{R},\bm{r})\cdot\bm{\sigma}_{\sigma\sigma'}\delta_{\tau\tau'} + \tfrac{1}{4}\vec{\bm{s}}(\bm{R},\bm{r})\cdot\bm{\sigma}_{\sigma\sigma'}\circ\vec{\tau}_{\tau\tau'},\end{aligned} \tag{116}$$

where $\bm{\sigma}$ and $\vec{\tau}$ are the spin (19) and isospin (27) Pauli matrices. The scalar-isoscalar $\rho_0(\bm{R},\bm{r})$, scalar-isovector $\vec{\rho}(\bm{R},\bm{r})$, vector-isoscalar $\bm{s}_0(\bm{R},\bm{r})$, and vector-isovector $\vec{\bm{s}}(\bm{R},\bm{r})$ densities can be obtained in a standard way by taking appropriate traces with the Pauli matrices. All necessary local densities can now be obtained by calculating at $r=0$ the derivatives in the total $\bm{\nabla} = \partial/\partial\bm{R}$ and relative $\bm{\partial} = \partial/\partial\bm{r}$ coordinates, up to the second order.

Without the proton-neutron mixing, which we neglect from now on in order to simplify the presentation, only the third components of isovectors are non-zero, and we can use the notation

$$\rho_1(\bm{R},\bm{r}) \equiv \vec{\rho}_3(\bm{R},\bm{r}) \quad\text{and}\quad s_1(\bm{R},\bm{r}) \equiv \vec{s}_3(\bm{R},\bm{r}). \tag{117}$$

The list of all required local densities then reads [39]:

$$\begin{aligned}\text{Matter:} \quad & \rho_t(\bm{R}) = \rho_t(\bm{R},0), & (118) \\ \text{Current:} \quad & \bm{j}_t(\bm{R}) = [\bm{k}\rho_t(\bm{R},\bm{r})]_{\bm{r}=0}, & (119) \\ \text{Kinetic:} \quad & \tau_t(\bm{R}) = [(\bm{k}^2 - \tfrac{1}{4}\bm{K}^2)\rho_t(\bm{R},\bm{r})]_{\bm{r}=0}, & (120) \\ \text{Spin:} \quad & \bm{s}_t(\bm{R}) = \bm{s}_t(\bm{R},0), & (121) \\ \text{Spin-current:} \quad & J^{ij}_t(\bm{R}) = [k^i s^j_t(\bm{R},\bm{r})]_{\bm{r}=0}, & (122) \\ \text{Spin-kinetic:} \quad & \bm{T}_t(\bm{R}) = [(\bm{k}^2 - \tfrac{1}{4}\bm{K}^2)\bm{s}_t(\bm{R},\bm{r})]_{\bm{r}=0}, & (123)\end{aligned}$$

where

$$\bm{k} = -i\bm{\partial} = -i\partial/\partial\bm{r} = \tfrac{1}{2i}\left(\bm{\nabla}^x - \bm{\nabla}^y\right) \quad\text{and}\quad \bm{K} = -i\bm{\nabla} = -i\partial/\partial\bm{R} = -i\left(\bm{\nabla}^x + \bm{\nabla}^y\right), \tag{124}$$

are momentum operators in the relative and total coordinate, and index $t=0,1$ distinguishes between the isoscalar and isovector components. The kinetic densities are usually defined in terms of the derivatives acting on the $\bm{x}$ and $\bm{y}$ coordinates (98), i.e., $(\bm{k}^2 - \tfrac{1}{4}\bm{K}^2) = \bm{\nabla}^x\cdot\bm{\nabla}^y$. There is also one density depending on $\bm{K}\otimes\bm{k}$ (tensor-kinetic density) [40, 38], which we do not discuss here because it appears only for tensor interactions. Since the Pauli matrices $\bm{\sigma}$ and momenta $\bm{k}$ are time-odd operators, wee see that densities $\rho_t(\bm{R})$, $\tau_t(\bm{R})$, and $J^{ij}_t(\bm{R})$ are time-even, and densities $\bm{j}_t(\bm{R})$, $\bm{s}_t(\bm{R})$, and $\bm{T}_t(\bm{R})$ are time-odd.

For an arbitrary central finite-range local potential with the full spin-isospin dependence [cf. the Gogny interaction in Eq. (63)],

$$G(\bm{x},\bm{y}) = W(\bm{x},\bm{y}) + B(\bm{x},\bm{y})P_\sigma - H(\bm{x},\bm{y})P_\tau - M(\bm{x},\bm{y})P_\sigma P_\tau, \tag{125}$$

we can now repeat the derivation of the LDA functional, by using expansions (100) and (107) in each spin-isospin channel. As a result, we obtain the interaction energy (direct and exchange terms combined) in the form

$$E^{\text{int}} = \sum_{t=0,1} \int d^3 \boldsymbol{R} \Big[ C_t^\rho \rho_t^2 + C_t^{\Delta\rho} \rho_t \Delta\rho_t + C_t^\tau (\rho_t \tau_t - \boldsymbol{j}_t^2) + C_t^s \boldsymbol{s}_t^2 + C_t^{\Delta s} \boldsymbol{s}_t \cdot \Delta \boldsymbol{s}_t + C_t^T (\boldsymbol{s}_t \cdot \boldsymbol{T}_t - \overset{\leftrightarrow 2}{J}_t) \Big], \quad (126)$$

where $\overset{\leftrightarrow 2}{J} = \sum_{ij} J^{ij} J_{ij}$. The energy density depends on six isoscalar and six isovector coupling constants that are simple moments of potentials, i.e.,

$$8 \begin{pmatrix} C_0^\rho \\ C_1^\rho \\ C_0^s \\ C_1^s \end{pmatrix} = \begin{pmatrix} 4 & -1 & 2 & -2 \\ 0 & -1 & 0 & -2 \\ 0 & -1 & 2 & 0 \\ 0 & -1 & 0 & 0 \end{pmatrix} \begin{pmatrix} W_0 + M_0' \\ M_0 + W_0' \\ B_0 + H_0' \\ H_0 + B_0' \end{pmatrix} \quad (127)$$

and

$$32 \begin{pmatrix} C_0^{\Delta\rho} \\ C_1^{\Delta\rho} \\ C_0^\tau \\ C_1^\tau \\ C_0^{\Delta s} \\ C_1^{\Delta s} \\ C_0^T \\ C_1^T \end{pmatrix} = \begin{pmatrix} 8 & -1 & 4 & -2 & -4 & 2 & -2 & 4 \\ 0 & -1 & 0 & -2 & -4 & 0 & -2 & 0 \\ 0 & 4 & 0 & 8 & 0 & -8 & 0 & -16 \\ 0 & 4 & 0 & 8 & 0 & 0 & 0 & 0 \\ 0 & -1 & 4 & 0 & 0 & 2 & -2 & 0 \\ 0 & -1 & 0 & 0 & 0 & 0 & -2 & 0 \\ 0 & 4 & 0 & 0 & 0 & -8 & 0 & 0 \\ 0 & 4 & 0 & 0 & 0 & 0 & 0 & 0 \end{pmatrix} \begin{pmatrix} W_2 \\ W_2' \\ B_2 \\ B_2' \\ H_2 \\ H_2' \\ M_2 \\ M_2' \end{pmatrix}, \quad (128)$$

where (for $X \equiv W$, $B$, $H$, or $M$)

$$X_0 = 4\pi \int dr \, r^2 X(r) \quad \text{and} \quad X_2 = \tfrac{4}{3}\pi \int dr \, r^4 X(r), \quad (129)$$

$$X_0' = 4\pi \int dr \, r^2 \pi_0^2(r) X(r) \quad \text{and} \quad X_2' = \tfrac{4}{3}\pi \int dr \, r^4 \pi_0(r)\pi_2(r) X(r). \quad (130)$$

Again we see, that for $\pi_0(r) = \pi_2(r) = 1$, the direct and exchange coupling constants are equal, $X_0' = X_0$ and $X_2' = X_2$, and hence only six coupling constants in energy density (126) are independent. This requires that the so-called time-odd coupling constants are linear combinations of the so-called time-even coupling constants [37]:

$$3 \begin{pmatrix} C_0^s \\ C_1^s \end{pmatrix} = \begin{pmatrix} -2 & -3 \\ -1 & 0 \end{pmatrix} \begin{pmatrix} C_0^\rho \\ C_1^\rho \end{pmatrix} \quad (131)$$

and

$$24 \begin{pmatrix} C_0^{\Delta s} \\ C_1^{\Delta s} \\ C_0^T \\ C_1^T \end{pmatrix} = \begin{pmatrix} -12 & -12 & 3 & 9 \\ -4 & -4 & 3 & -3 \\ 16 & 48 & -4 & 12 \\ 16 & -16 & 4 & -12 \end{pmatrix} \begin{pmatrix} C_0^{\Delta\rho} \\ C_1^{\Delta\rho} \\ C_0^\tau \\ C_1^\tau \end{pmatrix}. \quad (132)$$

It is well known that the local energy density (126) is also obtained for the Skyrme zero-range momentum-dependent interaction [41, 42, 43]. Without density-dependent and spin-orbit terms, this interaction reads

$$G(\boldsymbol{x}, \boldsymbol{y}) = t_0 (1 + x_0 P_\sigma) \, \delta(\boldsymbol{x} - \boldsymbol{y}) + \tfrac{1}{2} t_1 (1 + x_1 P_\sigma) \big[ \hat{\boldsymbol{k}}'^2 \, \delta(\boldsymbol{x} - \boldsymbol{y}) + \delta(\boldsymbol{x} - \boldsymbol{y}) \, \hat{\boldsymbol{k}}^2 \big]$$
$$+ t_2 (1 + x_2 P_\sigma) \, \hat{\boldsymbol{k}}' \cdot \delta(\boldsymbol{x} - \boldsymbol{y}) \, \hat{\boldsymbol{k}}, \quad (133)$$

where $\boldsymbol{k}' = i\boldsymbol{\partial}$ acts to the left, and $\boldsymbol{k} = -i\boldsymbol{\partial}$ acts to the right. For this interaction, the interaction energy has exactly the form given in Eq. (126), with coupling constants [39, 37] that depend on parameters $t_0$, $x_0$, $t_1$, $x_1$, $t_2$, and $x_2$,

$$8\begin{pmatrix} C_0^\rho \\ C_1^\rho \\ C_0^s \\ C_1^s \end{pmatrix} = \begin{pmatrix} -3 & 0 \\ -1 & -2 \\ -1 & 2 \\ -1 & 0 \end{pmatrix} \begin{pmatrix} t_0 \\ t_0 x_0 \end{pmatrix} \quad (134)$$

and

$$64 \begin{pmatrix} C_0^{\Delta\rho} \\ C_1^{\Delta\rho} \\ C_0^\tau \\ C_1^\tau \\ C_0^{\Delta s} \\ C_1^{\Delta s} \\ C_0^T \\ C_1^T \end{pmatrix} = \begin{pmatrix} -9 & 0 & 5 & 4 \\ 3 & 6 & 1 & 2 \\ 12 & 0 & 20 & 16 \\ -4 & -8 & 4 & 8 \\ 3 & -6 & 1 & 2 \\ 3 & 0 & 1 & 0 \\ -4 & 8 & 4 & 8 \\ -4 & 0 & 4 & 0 \end{pmatrix} \begin{pmatrix} t_1 \\ t_1 x_1 \\ t_2 \\ t_2 x_2 \end{pmatrix}. \quad (135)$$

For $\pi_0(r) = \pi_2(r) = 1$, the Skyrme interaction (133) exactly reproduces the LDA of the finite-range interaction (125), provided the Skyrme parameters are given by

$$t_0 = W_0 + M_0 \quad , \quad t_0 x_0 = B_0 + H_0, \quad (136)$$
$$t_1 = -W_2 - M_2 \quad , \quad t_1 x_1 = -B_2 - H_2, \quad (137)$$
$$t_2 = W_2 - M_2 \quad , \quad t_2 x_2 = B_2 - H_2. \quad (138)$$

Coupling constants of the Skyrme functional fulfill constraints (131) and (132). When the better approximation of the non-local density matrix is used, i.e., for $\pi_0(r) \neq 1$ or $\pi_2(r) \neq 1$ in Eq. (107), the Skyrme interaction cannot reproduce the LDA energy density. However, it is enough to release constraints (131) and (132), and treat all the twelve coupling constants as independent parameters, to recover the full freedom of the LDA local energy density.

Again we explicitly see that (exactly in the spirit of the effective field theory), the zero-range interaction can reproduce the same properties of nuclear systems as does the real effective interaction, provided the coupling constants in the energy density are either adjusted to data, or calculated from the real finite-range interaction. It is also clear that the zero-range interaction cannot be treated literally – it is significant only as a "generator" of the proper energy density, while all physical results depend only on this energy density, and not on the interaction itself. In particular, it is incorrect to look for exact eigenstates of the system interacting with the zero-range interaction; we know that for such an interaction the ground state does not exist because of the collapse. However, even for the finite-range effective interaction (for which the ground state does, in principle, exist) the exact ground state is irrelevant, because the interaction has been built to act only in the space of Slater determinants, see Sec. 4.2.

Of course, there is nothing magic or fundamental in the LDA to the energy density. It just reflects the fact that the nuclear one-body density matrix varies on a larger scale of distances than does the nuclear effective interaction. Validity of this approximation depends on the fundamental assumption that the total energy can be described as a functional of the one-body density matrix. The fact that we assumed a local effective interaction is not crucial – for non-local interactions the direct term becomes more complicated, but the LDA still holds [35]. However, effective interactions must, in fact, also depend on energy (Secs. 4.2 and 4.3), so the presented derivation of the LDA is not complete. One usually goes beyond the local energy density derived from approximations to one-body density, and one includes also terms that depend on local densities in a more complicated way, cf. the density-dependent term of the Gogny interaction (63).

Some people say: the LDA is just fitting of parameters – it is enough to have many parameters to fit anything one wants. This point of view simply disregards the success of the LDA in nuclear phenomenology. The effective field theory point of view is, in my opinion, more interesting, and potentially more fruitful. It regards the success of phenomenological LDA as indication that scales between quark-gluon QCD interactions and low-energy nuclear phenomena are indeed very well separated, and hence few numbers only are enough to define latter in terms of the former. The challenge of course remains: to look for derivations of these few numbers by decent fundamental theory, and to adjust these numbers to data and look for phenomena where the adjustments fail.

We finish this section by recalling the form of the HF equation (73), and that of the HF mean-field Hamiltonian (72), corresponding to the local-energy-density functional (126). Upon variation of the energy with respect to local densities, one obtains the HF equation (74) in spatial coordinates,

$$h_\alpha \psi_{i,\alpha}(\boldsymbol{r}\sigma) = \epsilon_{i,\alpha} \psi_{i,\alpha}(\boldsymbol{r}\sigma), \tag{139}$$

where $i$ numbers the neutron ($\alpha$=n) and proton ($\alpha$=p) orbitals, and

$$h_n = -\frac{\hbar^2}{2m}\Delta + \Gamma_0^{\text{even}} + \Gamma_0^{\text{odd}} + \Gamma_1^{\text{even}} + \Gamma_1^{\text{odd}}, \tag{140}$$

$$h_p = -\frac{\hbar^2}{2m}\Delta + \Gamma_0^{\text{even}} + \Gamma_0^{\text{odd}} - \Gamma_1^{\text{even}} - \Gamma_1^{\text{odd}}. \tag{141}$$

The isoscalar ($t$=0) and isovector ($t$=1) time-even and time-odd mean fields read

$$\Gamma_t^{\text{even}} = -\boldsymbol{\nabla}\cdot M_t(\boldsymbol{r})\boldsymbol{\nabla} + U_t(\boldsymbol{r}) + \tfrac{1}{2i}\left(\overleftrightarrow{\boldsymbol{\nabla}\sigma}\cdot\overleftrightarrow{B}_t(\boldsymbol{r}) + \overleftrightarrow{B}_t(\boldsymbol{r})\cdot\overleftrightarrow{\boldsymbol{\nabla}\sigma}\right), \tag{142}$$

$$\Gamma_t^{\text{odd}} = -\boldsymbol{\nabla}\cdot\left(\boldsymbol{\sigma}\cdot\boldsymbol{C}_t(\boldsymbol{r})\right)\boldsymbol{\nabla} + \boldsymbol{\sigma}\cdot\boldsymbol{\Sigma}_t(\boldsymbol{r}) + \tfrac{1}{2i}\left(\boldsymbol{\nabla}\cdot\boldsymbol{I}_t(\boldsymbol{r}) + \boldsymbol{I}_t(\boldsymbol{r})\cdot\boldsymbol{\nabla}\right), \tag{143}$$

where we defined the following mean-field potentials as functions of densities

$$U_t = 2C_t^\rho \rho_t + 2C_t^{\Delta\rho}\Delta\rho_t + C_t^\tau \tau_t, \tag{144}$$

$$\boldsymbol{\Sigma}_t = 2C_t^s \boldsymbol{s}_t + 2C_t^{\Delta s}\Delta\boldsymbol{s}_t + C_t^T \boldsymbol{T}_t, \tag{145}$$

$$M_t = C_t^\tau \rho_t,, \tag{146}$$

$$\boldsymbol{C}_t = C_t^T \boldsymbol{s}_t,, \tag{147}$$

$$\overleftrightarrow{B}_t = 2C_t^J \overleftrightarrow{J}_t, \tag{148}$$

$$\boldsymbol{I}_t = 2C_t^j \boldsymbol{j}_t. \tag{149}$$

Since neither in the effective interactions, (125) and (133), nor in the energy density (126), we showed the spin-orbit, tensor, or density-dependent terms, such contributions are not shown in the mean fields above. The mean-field Hamiltonian resulting from the LDA is simply given by local one-body potentials, with a complete dependence on spin, and by momentum-dependent terms that have the form of generalized effective-mass and spin-momentum couplings.